\documentclass[twocolumn,revtex4,apj,iop]{openjournal}
\usepackage{xcolor}
\usepackage{graphicx}
\usepackage{amsfonts}
\usepackage{amssymb}
\usepackage{url}
\usepackage[breaklinks,colorlinks,citecolor=blue,linkcolor=blue,urlcolor=blue]{hyperref}
\usepackage{amsmath}
\usepackage[varg]{txfonts}

\DeclareMathAlphabet{\mathbbold}{U}{bbold}{m}{n}

\renewcommand{\d}{\mathrm{d}}
\newcommand{\p}{\partial}

\newcommand{\calD}{\mathcal{D}}

\setcitestyle{numbers,square}
\begin{document}

\title{Backreaction in Numerical Relativity: Averaging on Newtonian gauge-like hypersurfaces in Einstein Toolkit cosmological simulations}

\author{Alexander Oestreicher\footnote{alexo@cp3.sdu.dk}}
\author{Sofie Marie Koksbang\footnote{koksbang@cp3.sdu.dk}}

\affiliation{CP3-Origins, University of Southern Denmark, \\ Campusvej 55, DK-5230 Odense M, Denmark\\}

\begin{abstract}
We introduce a spatial averaging scheme and use it to study the evolution of spatial averages in large-scale simulations of cosmological structure formation performed with the Einstein Toolkit. The averages are performed on the spatial hypersurfaces of the simulation setup which, at least initially, represent the hypersurfaces of statistical homogeneity and isotropy. We find only negligible cosmic backreaction on these hypersurfaces even on very small scales, but find significant curvature fluctuations of up to  $10\%$  in $\Omega_R$ for sub-volumes with radius $\sim200$ Mpc and even larger fluctuations in smaller sub-volumes. In addition, we quantify fluid flow in and out of these  sub-volumes. We find this to be significant, up to a $5\%$ change in the density between redshift $z=1$ and $z=0$ of a single sphere of radius $\sim200$ Mpc (and larger for smaller spheres). We suggest this may be important for studies basing averages on volumes co-moving with the simulation hypersurfaces.
\end{abstract}


\maketitle 
\tableofcontents

\section{Introduction}
Modern cosmology is based on the Friedmann-Lemaitre-Robertson-Walker (FLRW) solutions to Einstein's field equation. These models represent universes that are \emph{exactly} homogeneous and isotropic on their spatial hypersurfaces. The real universe is, however, only homogeneous and isotropic if averaged on scales above the homogeneity scale of roughly $100\;\mathrm{h}^{-1}\mathrm{Mpc}$ \cite{Yadav2010, Scrimgeour2012, Laurent2016, Goncalves2018, Goncalves2021}. Due to the non-linearity of Einstein's field equation, it is not given that the large-scale evolution of the Universe coincides with that of the FLRW models. Indeed, the averaged equations contain additional terms (so-called backreaction terms) as first demonstrated in \cite{Buchert2000}. The significance of this so-called cosmic backreaction has been heavily debated in the literature: Some authors for instance claim that backreaction is negligible in our universe \cite{Green2011, Green2012, Green2013, Green2014} while others suggest that backreaction could replace the need for dark energy entirely \cite{Buchert2000b, Schwarz2002, Rasanen2004}. The claims of \cite{Green2011, Green2012, Green2013, Green2014} have been disputed \cite{Buchert2015} and the averages used there have been shown to not represent a faithful description of the averaged dynamics of inhomogeneous spacetimes \cite{Clifton2019}. Nonetheless, it does seem increasingly unlikely that backreaction can explain away the entire need for dark energy. Indeed, such a scenario becomes decreasingly plausible as we obtain increasingly more ample and precise observations that fit well with the the $\Lambda$CDM model. However, reality is that observations do exhibit an array of tensions when interpreted within the $\Lambda$CDM model \cite{Abdalla2022}. These tensions could (partially) be due to backreaction. This is demonstrated in e.g. \cite{Bolejko2018, Kovacs2020, Heinesen2020, Clifton2024} which show that backreaction could in principle play an important role in terms of the Hubble tension. Thus, overall, in the era of precision cosmology, identifying if there are even percent level corrections to the evolution equations is important when seeking to interpret new and highly precise data.

One approach for learning about backreaction and its possible importance when interpreting observations is to consider inhomogeneous solutions to Einstein's field equation. This can be done using exact solutions to Einstein's equation but more recently it has also become possible to study this numerically, i.e. using general-relativistic cosmological simulations that either include relativistic effects or are based entirely on numerical relativity. These new tools e.g. include \texttt{gevolution} \cite{Adamek2016}, \texttt{GRAMSES} \cite{BarreraHinojosa2020, BarreraHinojosa2020b}, \texttt{CosmoGRAPH} \cite{Mertens2016} and cosmological simulations based on the \texttt{Einstein Toolkit} \cite{Macpherson2019}. These were all compared in \cite{Adamek2020} where they were found to agree well. The strength with using general-relativistic simulations rather than exact solutions to Einstein's equation is that the latter are only known for very simple inhomogeneities such as solutions representing a single void surrounded by an over-density. Such simple spacetimes cannot be expected to faithfully trace the evolution of the hierarchies of structures observed in the real universe. With general-relativistic simulations we can instead study the evolution of spacetimes with complicated structure formation based on Gaussian random fields.

The notion of averaging in cosmology is also related to the so-called fitting-problem discussed in e.g. chapter 16 of \cite{RelativisticCosmology_Ellis2012}.
The fitting-problem refers to the question of how spatial averages are related to observations. Astronomical observations are made on the lightcone and not spatial hypersurfaces. It is thus not clear which spatial hypersurfaces should be considered when averaging if the averages are to be sensibly related to observations -- if it is indeed possible to choose such a foliation. The fitting problem has been discussed in relation to backreaction studies where several different schemes for relating spatial averages to observations have been suggested in the literature (e.g \cite{Rasanen2010, Rasanen2009, Rosenthal2008, Larena2009}). In particular, we highlight the scheme proposed in \cite{Rasanen2010, Rasanen2009} where it was argued that spatial averages can be easily related to observations if the former are based on hypersurfaces of statistical homogeneity and isotropy, and assuming that light rays sample spacetime fairly (meaning that there are e.g. no opaque regions) with structures evolving slowly compared to the time it takes a light ray to traverse the assumed homogeneity scale. Under these requirements, spatial averages can be related to the mean redshift-distance relation according to
\begin{align}\label{eq:DA}
		H\frac{d}{d z}\left( (1+z )H\frac{dD_A }{d z } \right)& = -4\pi G\rho  D_A \\
		1+ z  &= 1/a,
\end{align}
where $z$ and $D_A$ here represent the {\em mean} redshift and angular diameter distance (i.e. the mean over many light rays in order to remove statistical fluctuations). The quantities $H$ and $a$ denote the averaged expansion rate and volume-scale factor (see Sect.~\ref{sec:2} for details on spatial averaging and the definition of the volume averaged scale factor). This relation has later been tested in various toy-models and exact (inhomogeneous) cosmological models \cite{Koksbang2020, Koksbang2019b, Koksbang2019}, demonstrating good agreement with the above relation if the requirements are fulfilled, but poor agreement when one or more of them are broken \cite{Koksbang2019c, Koksbang2020b} or when exotic features such as surface layers are introduced \cite{Lavinto2013}. We also note that the studies supporting the suggestions of \cite{Rasanen2010, Rasanen2009} are based on spacetimes where the 3-dimensional hypersurfaces of statistical homogeneteity and isotropy coincide with the hypersurfaces orthogonal to the fluid flow. We can therefore not rule out that the orthogonality to the fluid flow is necessary for \eqref{eq:DA} to be a good approximation. We nonetheless conclude that it is at least \emph{likely} that spatial averages are sensibly (\emph{viz.} related to at least some observables such as the redshift-distance relation) if the spatial hypersurfaces are those of statistical homogeneity and isotropy. 

In the following we therefore present a study of spatial averages and backreaction on such hypersurfaces of cosmological simulations performed with the \texttt{Einstein Toolkit}. We start by presenting our averaging scheme in Sect.~\ref{sec:2} and then explain the simulation setup and our post-processing of the simulation data in Sect.~\ref{sec:3}. In Sect.~\ref{sec:4} we present our results and compare them to results obtained earlier by others, before concluding in Sect.~\ref{sec:5}.

\section{Averaging Formalism}
\label{sec:2}
The onset of developing our averaging scheme is Buchert's averaging formalism originally introduced in \cite{Buchert2000}. Similar formalisms were later developed in \cite{Rasanen2009, Rasanen2010} in order to relate spatial averages to observations, leading to Eq.~\eqref{eq:DA}. These formalisms will be our starting point since our goal is to study the average evolution of the ET simulations, \emph{using an averaging scheme that is sensible for connecting averages to observations}. The numerical simulations we consider in this paper introduce a 3+1 decomposition of spacetime, slicing 4-dimensional spacetime into a family of 3-dimensional hypersurfaces $\Sigma$ of constant time $t$. Our goal is to calculate spatial averages on these surfaces as they (largely) represent the hypersurfaces of statistical homogeneity and isotropy (we discuss this point further in the beginning of Sect.~\ref{sec:3}). In order to compute the averages we must adapt the averaging formalism first introduced in \cite{Buchert2000} so that generic slicings of the spacetime and generic fluid flows can be considered. Such adaptations have been considered earlier in the literature, but not in the exact manner we require. For instance, \cite{Buchert2020} considers arbitrary slicings, but considers averages in spatial domains co-moving with the fluid flow. We, on the other hand, wish to consider spatial domains co-moving with the simulation hypersurfaces. Our formalism is formally most similar to that presented in \cite{Rasanen2010}, but again not exactly the same. We highlight the differences in detail in the following, but note here that the differences largely amount to defining different averaged quantities. 
Below, we introduce our averaging formalism. We begin with a subsection introducing the necessary basic definitions and decompositions. We then define our averaging scheme, keeping a general setting which we later specify to the particular setting of the simulations we consider.

\subsection{Basic Definitions and Decompositions}

We denote the vector normal to the hypersurfaces $\Sigma$ as $n^\alpha$ and assume it has been normalized such that $n_\alpha n^\alpha =-1$. The tensor 
\begin{align}
	h_{\alpha\beta} = g_{\alpha\beta} + n_\alpha n_\beta\; 
\end{align}
projects onto and is the metric on $\Sigma$. Here $g_{\alpha\beta}$ is the metric on the entire 4-dimensional spacetime. Additionally to the time $t$ constant on the spatial hypersurfaces, we introduce the proper time $s$ of an observer co-moving with the hypersurface. The derivative with respect to $s$ is given by
\begin{equation}
	\p_s = n^\alpha \nabla_\alpha\;, 
\end{equation}
and the derivative with respect to $t$ is 
\begin{equation}
	\p_t =\Gamma n^\alpha \nabla_\alpha\;, 
\end{equation}
where we defined 
\begin{equation}
	\Gamma \equiv \frac{\p s}{\p t}\;.
\end{equation}
As mentioned in \cite{Rasanen2010}, $\Gamma$ captures the time dilation due to the non-geodesic motion of the $n^\alpha$ frame.
We can define spatial covariant derivatives for scalars and vectors as 
\begin{align}
	\hat\nabla_\alpha f &= h_\alpha^\beta\nabla_\beta f \\
	\hat\nabla_\alpha f_\beta &= h_\alpha^\gamma h_\beta^\delta \nabla_\gamma f_\delta\;.
\end{align}
The spatial covariant derivative can be expressed using the 3-dimensional Christoffel symbols 
\begin{equation}
    {}^{(3)}\Gamma^\alpha_{\gamma\delta}= \frac{1}{2}h^{\alpha\beta}\left(\p_\gamma h_{\delta\beta}+\p_\delta h_{\gamma\beta}-\p_\beta h_{\gamma\delta}\right).
\end{equation}
With these, we can introduce the 3-dimensional Riemann tensor associated with the curvature on the hypersurface
\begin{equation}
    {}^{(3)}R^\alpha_{\beta\gamma\delta} = \p_\gamma {}^{(3)}\Gamma^\alpha_{\beta\delta}-\p_\delta {}^{(3)}\Gamma^\alpha_{\beta\gamma} + {}^{(3)}\Gamma^\alpha_{\gamma\lambda}{}^{(3)}\Gamma^\lambda_{\delta\beta}-{}^{(3)}\Gamma^\alpha_{\delta\lambda}{}^{(3)}\Gamma^\lambda_{\gamma\beta}
\end{equation}
and its contractions ${}^{(3)}R_{\alpha\beta}={}^{(3)}R^\gamma_{\alpha\gamma\beta}$ and ${}^{(3)}R = {}^{(3)}R^\alpha_\alpha$. We will refer to ${}^{(3)}R$ as the spatial curvature on $\Sigma$. We also introduce the following shorthand for the spatially projected traceless part of a rank-2 tensor
\begin{equation}
	f_{\langle\alpha\beta\rangle} = h_{(\alpha}^\gamma h_{\beta)}^\delta f_{\gamma\delta}-\frac{1}{3} h_{\alpha\beta}h^{\gamma\delta}f_{\gamma\delta}\;.
\end{equation}
We can write the co-variant derivative of the normal vector $n_\alpha$ as
\begin{equation}
	\nabla_\beta n_\alpha=\frac{1}{3} h_{\alpha \beta} \theta+\sigma_{\alpha \beta}-\dot n_\alpha n_\beta\;,
	\label{eq:decomb_nabla_n}
\end{equation}
where we define the volume expansion rate 
\begin{equation}
	\theta \equiv \nabla_\alpha n^\alpha=\hat{\nabla}_\alpha n^\alpha
\end{equation}
and the shear tensor 
\begin{equation}
	\sigma_{\alpha \beta} \equiv \nabla_{\langle\beta} n_{\alpha\rangle}=\hat{\nabla}_{\langle\beta} n_{\alpha\rangle}
\end{equation}
according to the usual conventions. We also define the shear scalar $\sigma^2 \equiv \frac{1}{2} \sigma_{\alpha \beta} \sigma^{\alpha \beta}$. Note that the vorticity term vanishes since $n^\alpha$ is hypersurface orthogonal. 
\newline\newline
We further introduce the extrinsic curvature defined as 
\begin{equation}
	K_{\alpha\beta} \equiv -h_\alpha^\gamma h_\beta^\delta\nabla_{(\gamma} n_{\delta)}=-h_\alpha^\gamma h_\beta^\delta\nabla_{\gamma} n_{\delta}\;,
\end{equation}
where the rounded brackets indicate symmetrization over the enclosed indices. The second equality holds since the antisymmetric part of this decomposition vanishes as $n^\alpha$ is hypersurface orthogonal. By definition, the extrinsic curvature tensor is equal to the negative expansion tensor 
\begin{equation}
	\theta_{\alpha\beta} \equiv h_\alpha^\gamma h_\beta^\delta\nabla_{(\gamma} n_{\delta)}\;.
\end{equation}
Both tensors are purely spatial and symmetric. The expansion rate $\theta$ introduced above is simply the trace of the expansion tensor $\theta = g^{\alpha\beta}\theta_{\alpha\beta}$. 
\newline\newline
We can decompose the energy-momentum tensor with respect to $n^\alpha$ as
\begin{equation}
	T_{\alpha\beta} = \rho^{(n)}n_\alpha n_\beta+p^{(n)}h_{\alpha\beta}+2q^{(n)}_{(\alpha}n_{\beta)}+\pi_{\alpha\beta}^{(n)}\;,
\end{equation}
where 
\begin{align}
	\rho^{(n)} &= n^\alpha n^\beta T_{\alpha\beta}\;, \\
	p^{(n)} &= h^{\alpha \beta} T_{\alpha\beta}\;, \\
	q^{(n)}_\alpha &= -h_\alpha^\beta n^\gamma n^T_{\beta\gamma}\;, \\
	\pi^{(n)}_{\alpha\beta} &= h_\alpha^\gamma h_\beta^\delta T_{\gamma\delta}-\frac{1}{3}h_{\alpha\beta}h^{\gamma\delta}T_{\gamma\delta} = T_{\langle \alpha\beta\rangle}\;, 
\end{align}
i.e. $\rho^{(n)}$ is the density in the n-frame, $p^{(n)}$ the pressure, $q^{(n)}_\alpha$ the energy flux and $\pi^{(n)}_{\alpha\beta}$ the anisotropic stress.
Similarly, we can decompose the energy momentum tensor with respect to the 4-velocity $u^{\alpha}$ as
\begin{equation}
	T_{\alpha\beta} = \rho^{(u)}u_\alpha u_\beta+p^{(u)}h^{(u)}_{\alpha\beta}+2q^{(u)}_{(\alpha}u_{\beta)}+\pi_{\alpha\beta}^{(u)}\;, 
\end{equation}
with $h^{(u)}_{\alpha\beta},\rho^{(u)}$, $p^{(u)}$, $q^{(u)}_\alpha$, $\pi^{(u)}_{\alpha\beta}$ defined analogously to the above n-frame quantities. Finally, we decompose the 4-velocity of an observer co-moving with the fluid $u^\alpha$ as 
\begin{equation}
	u^\alpha = \gamma \left(n^\alpha + v^\alpha\right)\;,
	\label{eq:u_decomposition}
\end{equation}
where $\gamma=-n_\alpha u^\alpha = (1-v^2)^{-1/2}$, with $v^2=v_\alpha v^\alpha$ and $v_\alpha n^\alpha = 0$.

\subsection{Average Evolution Equations}
We define the average of a scalar $f$ over a spatial domain $\calD$ lying on the hypersurfaces $\Sigma$ as
\begin{equation}
	\langle f\rangle (t) \equiv \frac{\int_\calD f \sqrt{h}\d^3 x}{\int_\calD \sqrt{h}\d^3 x}=\frac{1}{V_\calD}\int_\calD f \sqrt{h}\d^3 x\;,
    \label{eq:def_avg}
\end{equation}
where $h$ is the determinant of $h_{\alpha\beta}$. Taking time derivatives with respect to $t$ does not commute with spatial averages. Instead, we obtain the commutation relation 
\begin{equation}
	\p_t \langle f\rangle = \langle \p_t f\rangle + \langle \Gamma \theta f\rangle - \langle f\rangle \langle \Gamma\theta\rangle\;.
	\label{eq:commute_rel} 
\end{equation}

The expansion scalar $\theta$ gives the expansion rate of the local volume element with respect to the proper time $s$ \cite{Ehlers1961,Ehlers1993}. The quantity $\Gamma\theta$ gives the same with respect to the time $t$. We define the average scale factor $a$ such that 
\begin{equation}
	3\frac{\p_t a}{a}=\langle\Gamma\theta\rangle
	\label{eq:rel_a_theta}
\end{equation}
and 
\begin{equation}
	a\propto \left(V_\calD\right)^{1/3}\;,
\end{equation}
where an appropriate normalization may still be chosen. If $\Sigma$ is a hypersurface of statistical homogeneity and isotropy, this definition of $a$ is in line with the scale factor $a$ in equation \eqref{eq:DA} and hence we expect to approximately have $1+ z = 1/a$, where $z$ is the mean observed redshift (assuming that spacetime inhomogeneities evolve slowly compared to the time it takes a lightray to traverse the homogeneity scale, and spacetime is traced fairly by light rays) as discripted in \cite{Rasanen2010}.

Average evolution equations for $a(t)$ can be derived from the Raychaudhuri equation, the Hamiltonian constraint and the energy-momentum conservation equation 
\begin{align}
	\dot\theta + \frac{1}{3}\theta^2 =& -4\pi G_N (\rho^{(n)}+3p^{(n)})-2\sigma^2 \nonumber \\ 
	&+\dot n_\alpha \dot n^\alpha +\hat\nabla_\alpha \dot n^\alpha\;, \label{eq:raych_eq} \\
	\frac{1}{3}\theta^2 =&\; 8\pi G_N\rho^{(n)}-\frac{1}{2}{}^{(3)}R+\sigma^2\;, \label{eq:Ham_const}\\
	\dot \rho^{(n)}+\theta(\rho^{(n)}+p^{(n)})=&-\hat\nabla_\alpha q^{(n)\alpha}-2\dot n_\alpha q^{(n)\alpha}-\sigma_{\alpha\beta}\,\pi^{(n)\alpha\beta} \label{eq:mass_conservation}\;.
\end{align}
To do so it will be convenient to introduce the re-scaled variables 
\begin{align}
	\tilde \theta = \Gamma\theta\;, \quad \tilde \rho = \Gamma^2  \rho\;, \quad \tilde p = \Gamma^2 p \;, \nonumber \\ \quad \tilde \sigma = \Gamma \sigma\;, \quad {}^{(3)}\tilde R=\Gamma^2{}^{(3)}R\;,
 \label{eq:rescaled_variables}
\end{align}
similar to those introduced in \cite{Buchert2020} for the corresponding $u^\alpha$ quantities, since we have defined the scale factor such that $3\p_t a/a=\langle \tilde\theta\rangle$.  First multiplying \eqref{eq:raych_eq} and \eqref{eq:Ham_const} with a factor $\Gamma^2$ and \eqref{eq:mass_conservation} with a factor $\Gamma^3$, changing time derivatives with $\dot f = \Gamma^{-1} \p_t f$ and then averaging and using the two relations \eqref{eq:commute_rel} and \eqref{eq:rel_a_theta} leads to the set of averaged equations 
\begin{align}
	&3\frac{\p_t^2 a}{a} = -4\pi G_N \langle \tilde\rho^{(n)}+3\tilde p^{(n)}\rangle+Q+L\;, \label{eq:avg_friedm2} \\
	&3\left(\frac{\p_t a}{a}\right)^2 = 8\pi G_N\langle\tilde\rho^{(n)}\rangle-\frac{1}{2}\langle{}^{(3)}\tilde R\rangle-\frac{Q}{2}\;, \label{eq:avg_friedm1}\\
	\p_t&\langle\tilde\rho^{(n)}\rangle+3\frac{\p_t a}{a}\left(\langle\tilde\rho^{(n)}\rangle+\langle\tilde p^{(n)}\rangle\right) =\; \langle\tilde\theta\rangle\langle\tilde p\rangle-\langle\tilde\theta\tilde p^{(n)}\rangle \nonumber \\
	& + \langle\rho^{(n)}\p_t\Gamma^2\rangle -\langle\Gamma\hat\nabla_\alpha (\Gamma^2 q^{(n)\alpha})\rangle-\langle\Gamma^3\sigma_{\alpha\beta}\pi^{(n)\alpha\beta}\rangle\;. 
	\label{eq:avg_energy_momentum_consv}
\end{align}
In the final step of deriving \eqref{eq:avg_energy_momentum_consv} we used that $\dot n_\alpha = \Gamma^{-1}\hat\nabla_\alpha\Gamma$ and added a term $3\p_t a/a \langle\tilde p\rangle - \langle\tilde\theta\rangle\langle\tilde p\rangle = 0$ to recover the usual shape of this equation. 
We have defined the two backreaction terms 
\begin{align}
	Q &= \frac{2}{3}\left( \langle\tilde\theta^2\rangle-\langle\tilde\theta\rangle^2\right)-2\langle\tilde\sigma^2\rangle\;, \\
	L &= \langle\p_t n_\alpha \p_t n^\alpha\rangle +\langle\Gamma^2\hat\nabla_\alpha \dot n^\alpha\rangle+\langle\theta\p_t\Gamma\rangle\;.
\end{align}
Using again that $\dot n_\alpha = \Gamma^{-1}\hat\nabla_\alpha\Gamma$, we can rewrite $L$ as 
\begin{align}
	L= \langle g^{\alpha\beta}\Gamma\hat\nabla_\alpha\hat\nabla_\beta \Gamma\rangle +\langle\theta\p_t\Gamma\rangle\;.
	\label{eq:L}
\end{align}
The set of equations \eqref{eq:avg_friedm2}, \eqref{eq:avg_friedm1} and \eqref{eq:avg_energy_momentum_consv} resemble the two Friedmann equations and the energy-momentum conservation equation, but contain further terms. In the usual FLRW scenario the two Friedmann equations can be combined to yield the energy-momentum conservation equation. Here, the equations are independent and \eqref{eq:avg_friedm1} and \eqref{eq:avg_friedm2} can instead be combined to yield
\begin{align}
	8\pi G_N &\left(\p_t\langle\tilde\rho^{(n)}\rangle+3\frac{\p_t a}{a}\left(\langle\tilde\rho^{(n)}\rangle+\langle\tilde p^{(n)}\rangle\right) \right) \nonumber \\
	&=\frac{1}{2}\left(\frac{\p_t (a^2\langle {}^{(3)}\tilde R\rangle)}{a^2}+\frac{\p_t (a^6 Q)}{a^6}\right)+2\frac{\p_t a}{a}L\;.
\end{align}
The set of equations \eqref{eq:avg_friedm2}, \eqref{eq:avg_friedm1} and \eqref{eq:avg_energy_momentum_consv} are formally equivalent to the ones derived in \cite{Rasanen2010}, but \cite{Rasanen2010} derives them for the non-rescaled variables rather than the re-scaled variables \eqref{eq:rescaled_variables} we introduced. The re-scaled and non-rescaled variables are equivalent if there is no difference between the proper time of an observer co-moving with the hypersurfaces $s$ and the time constant on the hypersurfaces $t$. This is the case as long as motion is non-relativistic and there are no strong gravitational fields, which we expect to be the case for the simulations we study in this paper. A deeper discussion of conditions for this can be found in \cite{Rasanen2010}.

\section{Numerical Simulations}\label{sec:3}
The simulation data we study was kindly provided by Hayley J. Macpherson. The same data has earlier been used in \cite{Macpherson2023, Macpherson2024, Koksbang2024} and similar data created with the same code and the same initial conditions generator with similar power spectra was used in e.g. \cite{Macpherson2017,Macpherson2019,Macpherson2021,Williams2024}. The data was created using the \texttt{Einstein Toolkit}\footnote{\url{einsteintoolkit.org}} (ET) \cite{ET2023}, which is a community driven software platform for solving Einstein's equation numerically. 

The simulations are set up as a linearly perturbed Einstein-de Sitter universe, meaning they only contain matter, with curvature set to zero initially. The Hubble parameter is set to $H_0=100\;\mathrm{h}\; \mathrm{km}\mathrm{s}^{-1}\mathrm{Mpc}^{-1}$, with $\mathrm{h}=0.45$. This choice results in a universe with age $14.5\;\mathrm{Gyr}$, i.e. similar to the age of our own universe. The cosmological fluid is further assumed to be a pressureless perfect fluid, which is implemented as $p^{(u)}\ll \rho^{(u)}$ (as discussed in \cite{Macpherson2019}, the pressure can for technical reasons not be set to vanish identically). To high precision, the fluid energy-momentum tensor is thus simply
\begin{equation}
    T_{\mu\nu} = \rho^{(u)}u_\mu u_\nu\;.
    \label{eq:T_u}
\end{equation}
The initial conditions are set up using the thorn \texttt{FLRWSolver}\footnote{\url{https://github.com/hayleyjm/FLRWSolver_public}} presented in \cite{Macpherson2017} and then evolved using the thorns \texttt{ML\_BSSN} \cite{Brown2009}, evolving the spacetime variables, and \texttt{GRHydro} \cite{Mosta2014}, evolving the hydrodynamics. 

\texttt{FLRWSolver} sets up initial conditions for the simulation as Gaussian Random fields based on matter power spectra generated by e.g. \texttt{CLASS}\footnote{\url{http://class-code.net/}}, similar to methods used for setting initial conditions in standard Newtonian N-body simulations. All power below scales corresponding to $\sim $ 10 grid cells of a simulation was removed in the considered simulations in order to minimize numerical error with under-sampling small scale modes.  Any parameters not mentioned above were set in accordance with standard Planck parameters.

The simulation assumes the ADM metric with the shift set to zero 
\begin{equation}
	\d s^2 = -\alpha^2\; \d t^2 + \gamma_{ij}\; \d x_i \d x_j
	\label{eq:adm_metric}
\end{equation}
and evolves the initial conditions from $z = 1000$ to $z \simeq 0$, where $z$ is the redshift associated with the background EdS model. The initial conditions are set using the linearly perturbed FLRW metric in the longitudinal gauge, considering only scalar perturbations
\begin{equation}
	\d s^2 = - a^2(\eta)(1+2\psi)\d\eta^2+a^2(\eta)(1-2\phi)\delta_{ij}\d x^i\d x^j\;,
    \label{eq:lin_pert_metric}
\end{equation}
where $\eta$ is the conformal time and $\psi,\phi$ are the Bardeen potentials. The quantities $\alpha$ and $\gamma_{ij}$ are chosen such that \eqref{eq:adm_metric} matches \eqref{eq:lin_pert_metric} initially and $\psi$ is set equal to $\phi$. This means that our initial hypersurface is in the Newtonian Gauge and as long as the perturbations remain in the linear regime we will stay within this gauge. Since initial conditions are set up as Gaussian random fields in this gauge, we expect the corresponding spatial hypersurfaces to represent those of statistical homogeneity and isotropy, at least while in the linear regime. We therefore choose these to compute averages on these hypersurfaces, i.e. on the spatial hypersurfaces of our simulation. With this choice, the time parameter constant on the hypersurfaces, $t$, is conformal time $\eta$.

\subsection{Post-Processing}
The ET simulations give $\alpha, \gamma_{ij}, K_{ij}, \rho^{(u)}, \vec v$ as outputs. In order to analyse the output and calculate spatial averages we use \texttt{mescaline} \cite{Macpherson2019} a post-processing software specifically written to evaluate numerical simulations of the universe created with the ET. 

\texttt{mescaline} calculates a number of useful quantities from the outputs of the ET, including e.g. the 3-dimensional Christoffel symbols ${}^{(3)}\Gamma^\alpha_{\beta\gamma}$ and the spatial curvature ${}^{(3)}R$. \texttt{mescaline} also implements the fluid intrinsic averaging formalism suggested in \cite{Buchert2020}, which considers averages in the frame co-moving with the fluid (with 4-velocity $u^\alpha$). 

As our goal is to calculate averages on the simulation hypersurfaces with the averaging formalism presented in Sect.~\ref{sec:2}, we introduce this averaging formalism into \texttt{mescaline}. We therefore need all the terms in the three equations \eqref{eq:avg_friedm1}, \eqref{eq:avg_friedm2} and \eqref{eq:avg_energy_momentum_consv} in terms of the ET simulation output and quantities available through \texttt{mescaline}. To calculate the necessary terms, we first note that with the choice of the metric \eqref{eq:adm_metric}, the normal vector is given by 
\begin{equation}
	n_\alpha = (-\alpha,0,0,0)\;, \qquad n^\alpha = (\alpha^{-1},0,0,0)
\end{equation}
and 
\begin{equation}
	\Gamma = \alpha\;.
\end{equation}
The spatial metric on the hypersurface and its inverse are simply
\begin{equation}
	h_{\mu\nu} = 
	\begin{pmatrix}
		0 & \vec 0  \\
		\vec 0 & \gamma_{ij} 
	\end{pmatrix}\;,
	\qquad 
	h^{\mu\nu} = 
	\begin{pmatrix}
		0 & \vec 0  \\
		\vec 0 & \gamma_{ij}^{-1}
	\end{pmatrix}
\end{equation}
and the projection tensor $h^\mu_\nu$ is given by 
\begin{equation}
	h^\mu_\nu  = \delta^\mu_\nu + n^\mu n_\nu = \begin{pmatrix}
		0 & \vec 0 \\
		\vec 0 & \mathbbold{1}_3
	\end{pmatrix}\;.
\end{equation}
The expansion scalar and shear are by definition 
\begin{align}
	\theta = -K = \gamma^{ij}K_{ij}\;,
\end{align}
and 
\begin{equation}
	\sigma_{ij}= -K_{ij}-\frac{1}{3}\gamma_{ij}\theta\;, 
\end{equation}
which we can directly calculate from the ET outputs read in by \texttt{mescaline}. From this we can easily calculate $\sigma$ and thereafter $Q$. In order to calculate the second backreaction term, $L$, we first use that $\Gamma = \alpha$, leaving us with 
\begin{align}
	L &= \langle g^{\alpha\beta}\alpha\hat\nabla_\alpha\hat\nabla_\beta \alpha\rangle +\langle\theta\p_t\alpha\rangle \nonumber \\
    &= \langle \alpha \gamma^{ij} \p_i\p_j \alpha\rangle +\langle\theta\p_t\alpha\rangle\;,
\end{align}
where we used that $\hat \nabla$ is a spatial derivative and $\alpha$ a scalar. We also need the fluid variables in the normal frame $\rho^{(n)}$, $p^{(n)}$, $q^{(n)}_\alpha$, $\pi^{(n)}_{\alpha\beta}$. These can be calculated from suitable projections of \eqref{eq:T_u}. For the density we have 
\begin{equation}
	\rho^{(n)}=n^\alpha n^\beta T_{\alpha\beta}=n^\alpha n^\beta u_\alpha u_\beta \rho^{(u)}=\gamma^2\rho^{(u)}\;,
\end{equation}
where we used that $u_\alpha n^\alpha = -\gamma$. For the pressure we find
\begin{align}
p^{(n)}=h^{\alpha\beta}T_{\alpha\beta}=h^{\alpha\beta}u_\alpha u_\beta \rho^{(u)}= u_i u^i \rho^{(u)} = \gamma^2 v_i v^i \rho^{(u)}\;,
\end{align}
where we used that $u_i=\gamma v_i$, which can be seen from \eqref{eq:u_decomposition} and by the fact that the spatial components of $n_\alpha$ are zero. For the energy flux we find 
\begin{align}
	q^{(n)}_\alpha = -h_\alpha^\beta n^\gamma T_{\beta\gamma}=-h_\alpha^\beta n^\gamma u_\beta u_\gamma \rho^{(u)} =  h_\alpha^\beta u_\beta \gamma\rho^{(u)}\;, \nonumber
\end{align}
leaving us with 
\begin{align}
q^{(n)}_0= 0 \quad \mathrm{and} \quad q^{(n)}_i = u_i \gamma\rho^{(u)} = v_i \gamma^2\rho^{(u)}\;.
\end{align}
Lastly, for the anisotropic stress we find
\begin{align}
	\pi^{(n)}_{\alpha\beta} &= h_\alpha^\gamma h_\beta^\delta T_{\gamma\delta}-\frac{1}{3}h_{\alpha\beta}h^{\gamma\delta}T_{\gamma\delta} \nonumber \\ 
	&= \left( h_\alpha^\gamma h_\beta^\delta u_\gamma u_\delta-\frac{1}{3}h_{\alpha\beta}h^{\gamma\delta}u_\gamma u_\delta\right)\rho^{(u)}\;, \nonumber
\end{align}
which gives us 
\begin{align}
	\pi^{(n)}_{00}=\pi^{(n)}_{0i}=0 \quad \mathrm{and} \quad  \pi^{(n)}_{ij}&=\left( u_i u_j-\frac{1}{3}h_{ij}u^k u_k\right)\rho^{(u)} \nonumber \\
	&=\left( v_i v_j-\frac{1}{3}h_{ij}v^k v_k\right)\gamma^2\rho^{(u)}\;.
\end{align}
With this we now have all the terms appearing in the three equations \eqref{eq:avg_friedm1}, \eqref{eq:avg_friedm2} and \eqref{eq:avg_energy_momentum_consv}. Most of the above terms can be computed directly through quantities already available through \texttt{mescaline.} The main exception is $\langle\Gamma\hat\nabla_\mu (\Gamma^2 q^{(n)\mu})\rangle$ appearing in \eqref{eq:avg_energy_momentum_consv}, where we need to evaluate the covariant derivative. We find
\begin{align}
	\Gamma\hat\nabla_\mu (\Gamma^2 q^{(n)\mu}) &= \alpha\hat\nabla_\mu (\alpha^2 q^{(n)\mu}) \nonumber \\
	&=  2\alpha^2 q^{(n)\mu}\hat\nabla_\mu \alpha  + \alpha^3\hat\nabla_\mu q^{(n)\mu} \nonumber \\ 
	&= 2\alpha^2 q^{(n)i}\p_i \alpha  + \alpha^3\p_i q^{(n)i} + \alpha^3{}^{(3)}\Gamma^i_{ij} q^{(n)j}\;, 
\end{align}
where we again used that $\hat\nabla$ is the spatial covariant derivative and therefore has no time component, and that the energy flux is purely spatial. 
\newline

\texttt{mescaline} calculates the volume average \eqref{eq:def_avg} in the following manner
\begin{equation}
	\langle f\rangle = \frac{\sum_i f(x_i,y_i,z_i)\sqrt{h(x_i,y_i,z_i)}(\Delta x)^3}{\sum_i \sqrt{h(x_i,y_i,z_i)}(\Delta x)^3},
\end{equation}
with the sum including all points $i$ that are within the user-defined sphere and therefore fulfill the condition
\begin{equation}
	r_D^2 \geq \left(x_i - x_{origin}\right)^2+\left(y_i - y_{origin}\right)^2+\left(z_i - z_{origin}\right)^2\;.
\end{equation}
Here, $\Delta x$ is the side length of an individual grid cell. The scale factor is calculated from the volume as 
\begin{equation}
    a = \left(\sum_i \sqrt{h(x_i,y_i,z_i)}(\Delta x)^3\right)^{1/3}
\end{equation}
and normalized such that it would be one today if the sphere expanded exactly as the average of the entire simulation box. The Hubble rate $\p_t a/a$ is calculated from $\langle \Gamma\theta\rangle$ via the relation \eqref{eq:rel_a_theta}.

\begin{figure*}[]
	\centering
	\includegraphics[width=0.33\linewidth]{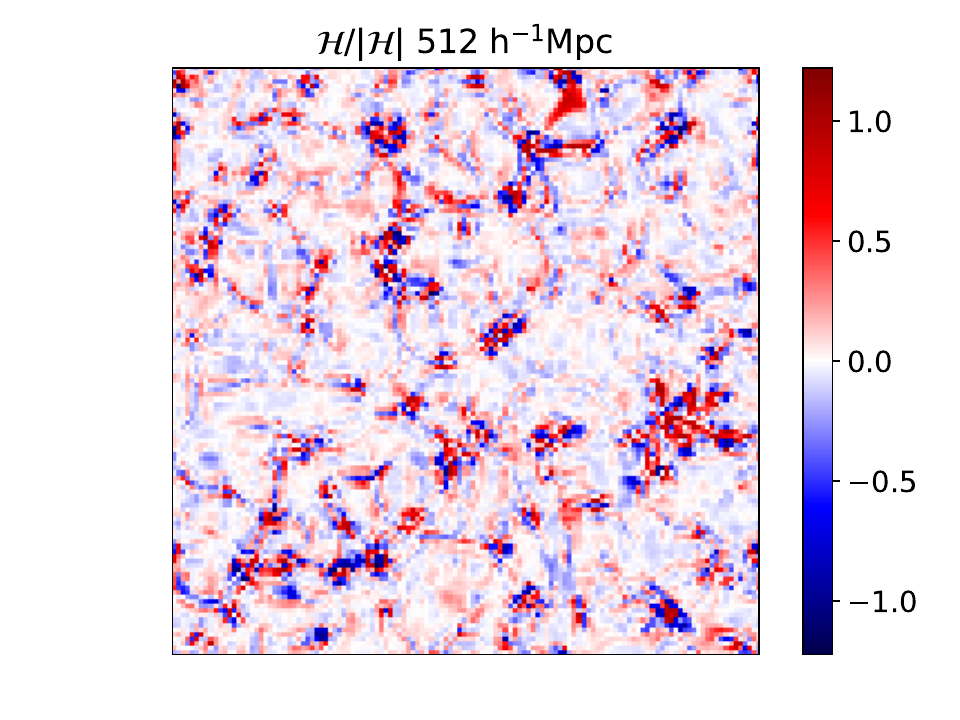}%
	\includegraphics[width=0.33\linewidth]{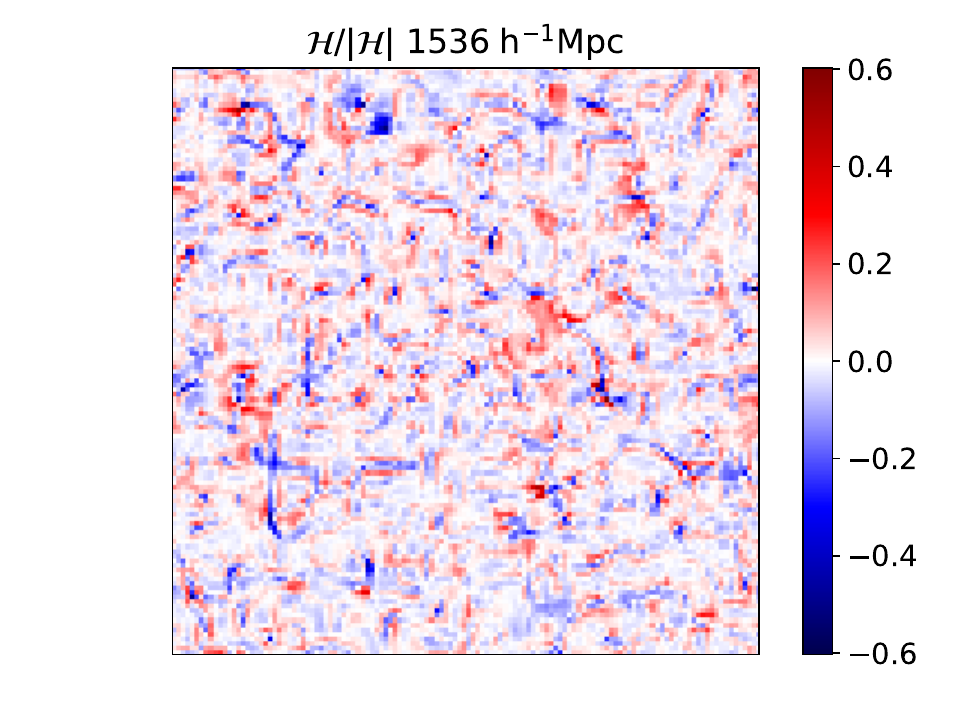}%
	\includegraphics[width=0.33\linewidth]{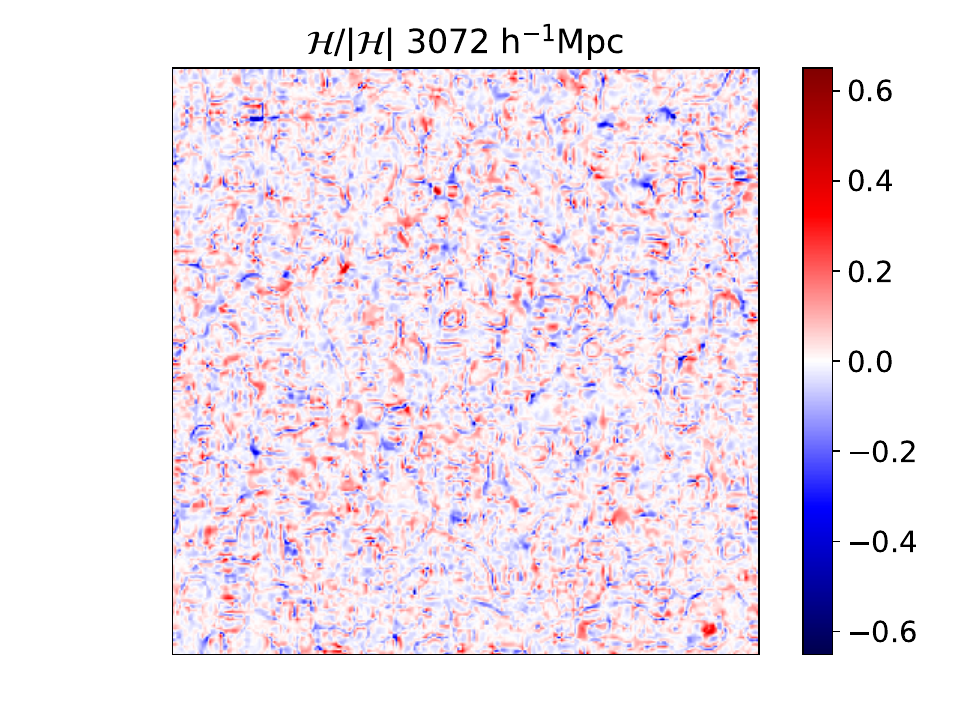}
	\includegraphics[width=0.33\linewidth]{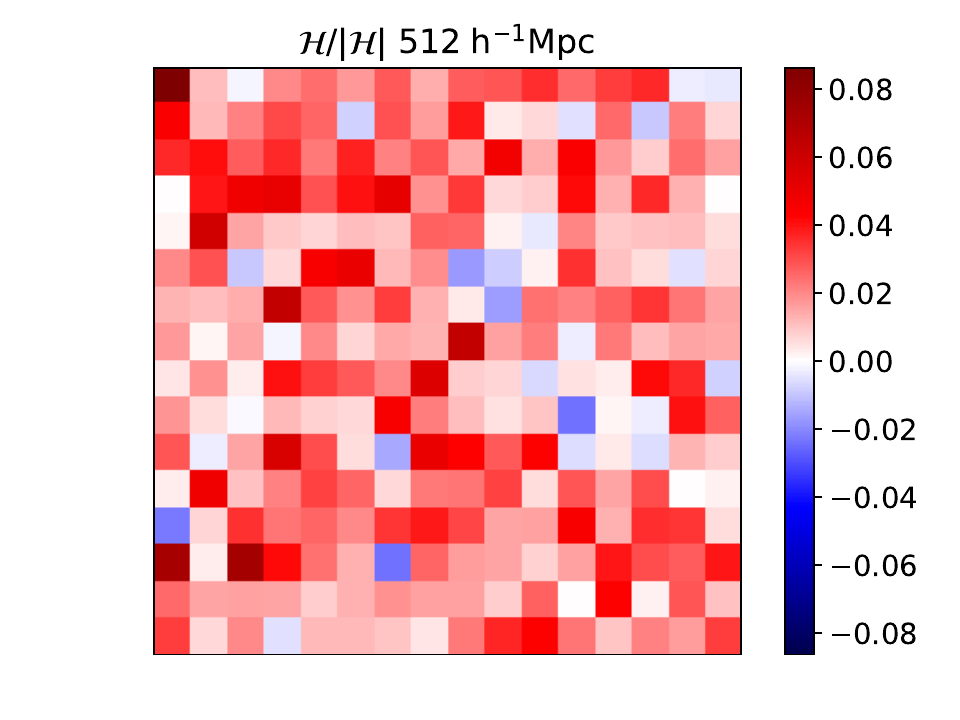}%
	\includegraphics[width=0.33\linewidth]{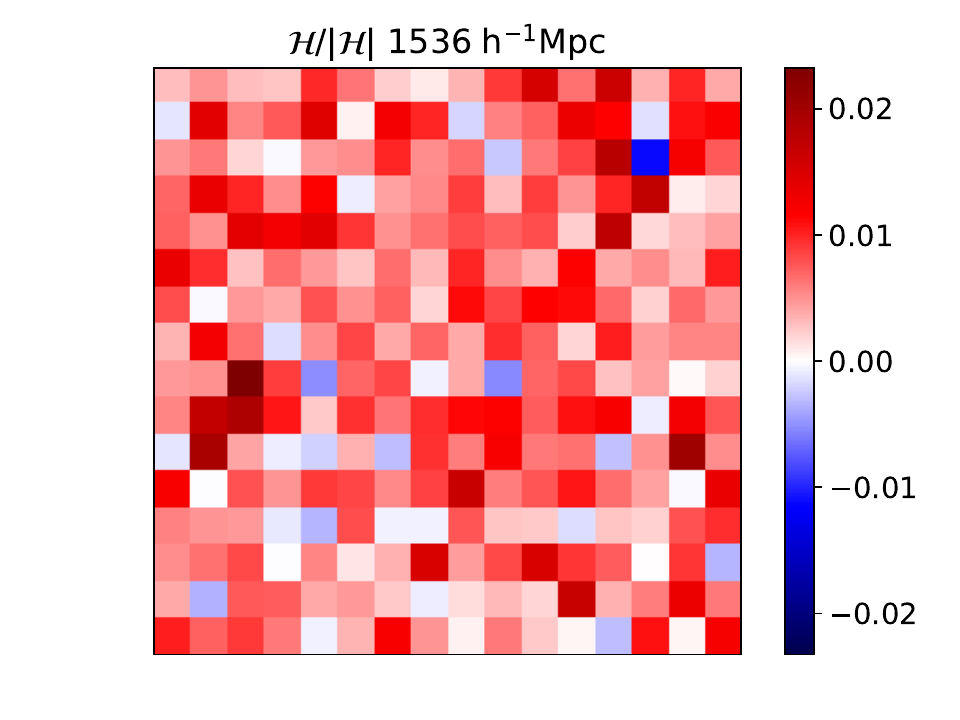}%
	\includegraphics[width=0.33\linewidth]{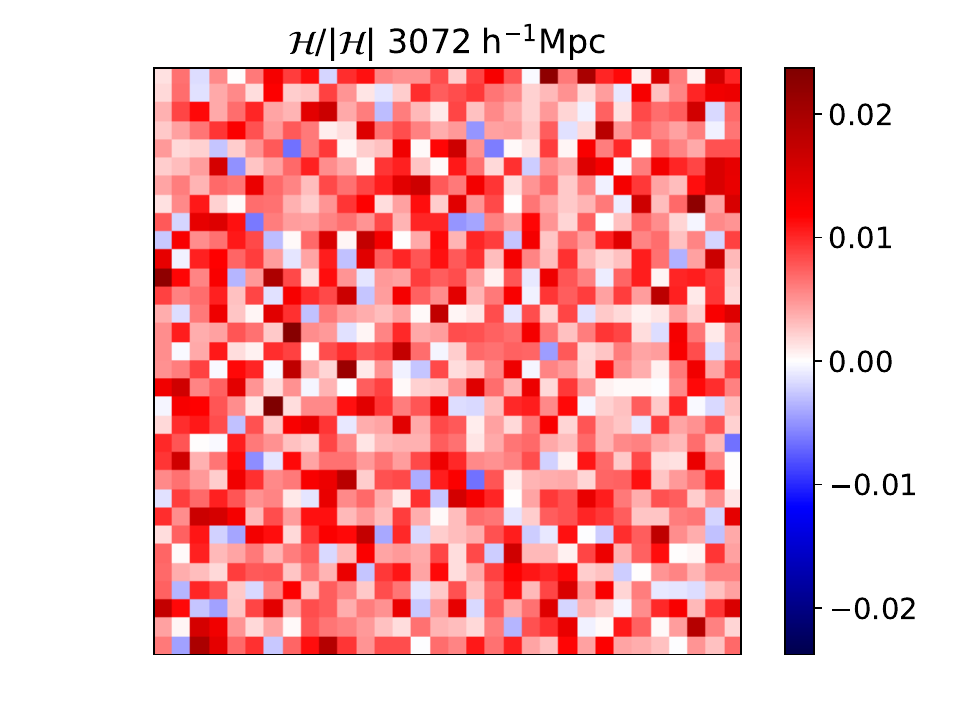}
	\caption{Constraint violation in the different simulations studied in this paper. The bottom row shows the violation on a coarse grained grid, where each grid point corresponds to smoothing over $8\times 8\times 8$ of the original simulation cells.}
    \label{Fig:constraint_violation}
\end{figure*}

\subsubsection{Conversion to Cosmic Time}
As mentioned earlier, the ET simulations are initialized with the metric \eqref{eq:lin_pert_metric}, thereby choosing the time parameter constant on the hypersurfaces to be the conformal time, $\eta$, and the initial lapse $\alpha_\mathrm{ini}=a\sqrt{(1-2\psi)}$. Since $\Gamma = \alpha$ and we introduced the re-scaled variables \eqref{eq:rescaled_variables} using $\Gamma$, this means that our re-scaled variables are multiplied by powers of the scale factor. We would like to avoid this and therefore introduce $\hat \alpha$ and $\hat \Gamma$ such that 
\begin{equation}
    \alpha = \bar a \hat \alpha = \Gamma = \bar a \hat \Gamma\;,
\end{equation}
where $\bar a$ is the average scale factor of the entire simulation box. We also introduce re-scaled variables $\hat \rho$ etc., defined with $\hat \Gamma$. With these replacements, the three equations \eqref{eq:avg_friedm2},\eqref{eq:avg_friedm1} and \eqref{eq:avg_energy_momentum_consv} become 
\begin{align}
	&3\frac{1}{\bar a^2}\left(\frac{\p_t^2 a}{a} - \frac{\p_t a}{a}\frac{\p_t \bar a}{\bar a}\right) = -4\pi G_N \langle \hat\rho^{(n)}+3\hat p^{(n)}\rangle+\hat Q+\hat L\;, \label{eq:avg_friedmann2_conformal} \\
	&3\frac{1}{\bar a^2}\left(\frac{\p_t a}{a}\right)^2 = 8\pi G_N\langle\hat\rho^{(n)}\rangle-\frac{1}{2}\langle{}^{(3)}\hat R\rangle-\frac{\hat Q}{2}\;, \label{eq:avg_friedmann1_conformal} \\
	\p_t&\langle\hat\rho^{(n)}\rangle+3\frac{\p_t a}{a}\left(\langle\hat\rho^{(n)}\rangle+\langle\hat p^{(n)}\rangle\right) =\; \bar a \langle\hat\theta\rangle\langle\hat p\rangle-\bar a\langle\hat\theta\hat p^{(n)}\rangle \nonumber \\
	& + \langle\rho^{(n)}\p_t\hat\Gamma^2\rangle -\bar a\langle\hat\Gamma\hat\nabla_\alpha (\hat\Gamma^2 q^{(n)\alpha})\rangle-\bar a\langle\hat\Gamma^3\sigma_{\alpha\beta}\pi^{(n)\alpha\beta}\rangle\;,
    \label{eq:avg_energy_consv_conformal} 
\end{align}
where we defined 
\begin{align}
	\hat Q &= \frac{2}{3}\left( \langle\hat\theta^2\rangle-\langle\hat\theta\rangle^2\right)-2\langle\hat\sigma^2\rangle\;, \\
	\hat L &= \langle g^{\alpha\beta}\hat \Gamma\hat\nabla_\alpha\hat\nabla_\beta \hat \Gamma\rangle +\bar a^{-1}\langle\theta\p_t\hat \Gamma\rangle\;.
\end{align}
The additional term in \eqref{eq:avg_friedmann2_conformal} appears due to the time derivative of $\Gamma$ appearing in $L$. Note that 
\begin{equation}
    L = \hat a^2\hat L + 3\frac{\p_t a}{a}\frac{\p_t \bar a}{\bar a}\;.
\end{equation}
The set of equations \eqref{eq:avg_friedmann2_conformal}, \eqref{eq:avg_friedmann1_conformal} and  \eqref{eq:avg_energy_consv_conformal} now resemble the Friedmann equations in conformal time. If we introduce the time coordinate $\tau$ via the relation $\bar a \d t = \d\tau$ we are left with the following set of equations
\begin{align}
	&3\frac{\p_\tau^2 a}{a} = -4\pi G_N \langle \hat\rho^{(n)}+3\hat p^{(n)}\rangle+\hat Q+\hat L\;,  \label{eq:avg_friedm2_v2} \\
	&3\left(\frac{\p_\tau a}{a}\right)^2 = 8\pi G_N\langle\hat\rho^{(n)}\rangle-\frac{1}{2}\langle{}^{(3)}\hat R\rangle-\frac{\hat Q}{2}\;,  \label{eq:avg_friedm1_v2}\\
	\p_\tau&\langle\hat\rho^{(n)}\rangle+3\frac{\p_\tau a}{a}\left(\langle\hat\rho^{(n)}\rangle+\langle\hat p^{(n)}\rangle\right) =\; \langle\hat\theta\rangle\langle\hat p\rangle-\langle\hat\theta\hat p^{(n)}\rangle \nonumber  \\
	& + \langle\rho^{(n)}\p_\tau\hat\Gamma^2\rangle -\langle\hat\Gamma\hat\nabla_\alpha (\hat\Gamma^2 q^{(n)\alpha})\rangle-\langle\hat\Gamma^3\sigma_{\alpha\beta}\pi^{(n)\alpha\beta}\rangle\;,
    \label{eq:avg_energy_consv_v2}
\end{align}
which is now the exact same as the set of equations \eqref{eq:avg_friedm2}, \eqref{eq:avg_friedm1} and \eqref{eq:avg_energy_momentum_consv} only with a redefined time coordinate and lapse. Since $t$ corresponded to conformal time, $\tau$ corresponds to cosmic time. The equations \eqref{eq:avg_friedm2_v2} and \eqref{eq:avg_friedm1_v2} can again be combined to yield
\begin{align}
	8\pi G_N &\left(\p_\tau\langle\hat\rho^{(n)}\rangle+3\frac{\p_\tau a}{a}\left(\langle\hat\rho^{(n)}\rangle+\langle\hat p^{(n)}\rangle\right) \right) \nonumber \\
	&=\frac{1}{2}\left(\frac{\p_\tau (a^2\langle {}^{(3)}\hat R\rangle)}{a^2}+\frac{\p_\tau (a^6 \hat Q)}{a^6}\right)+2\frac{\p_\tau a}{a}\hat L\;.
    \label{eq:LRQC_rel}
\end{align}
\begin{figure*}[]
	\centering
	\includegraphics[width=0.33\linewidth]{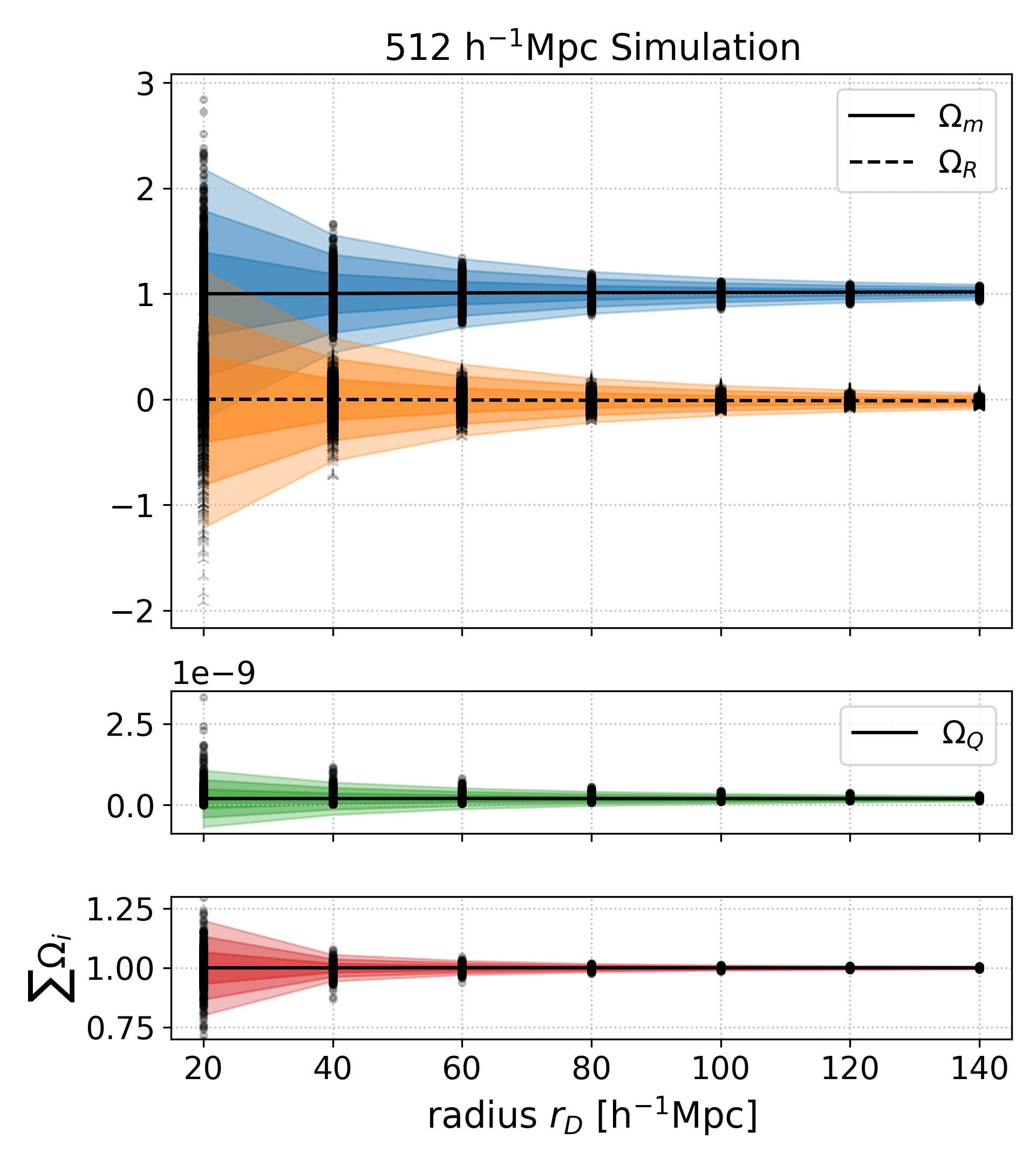}%
	\includegraphics[width=0.33\linewidth]{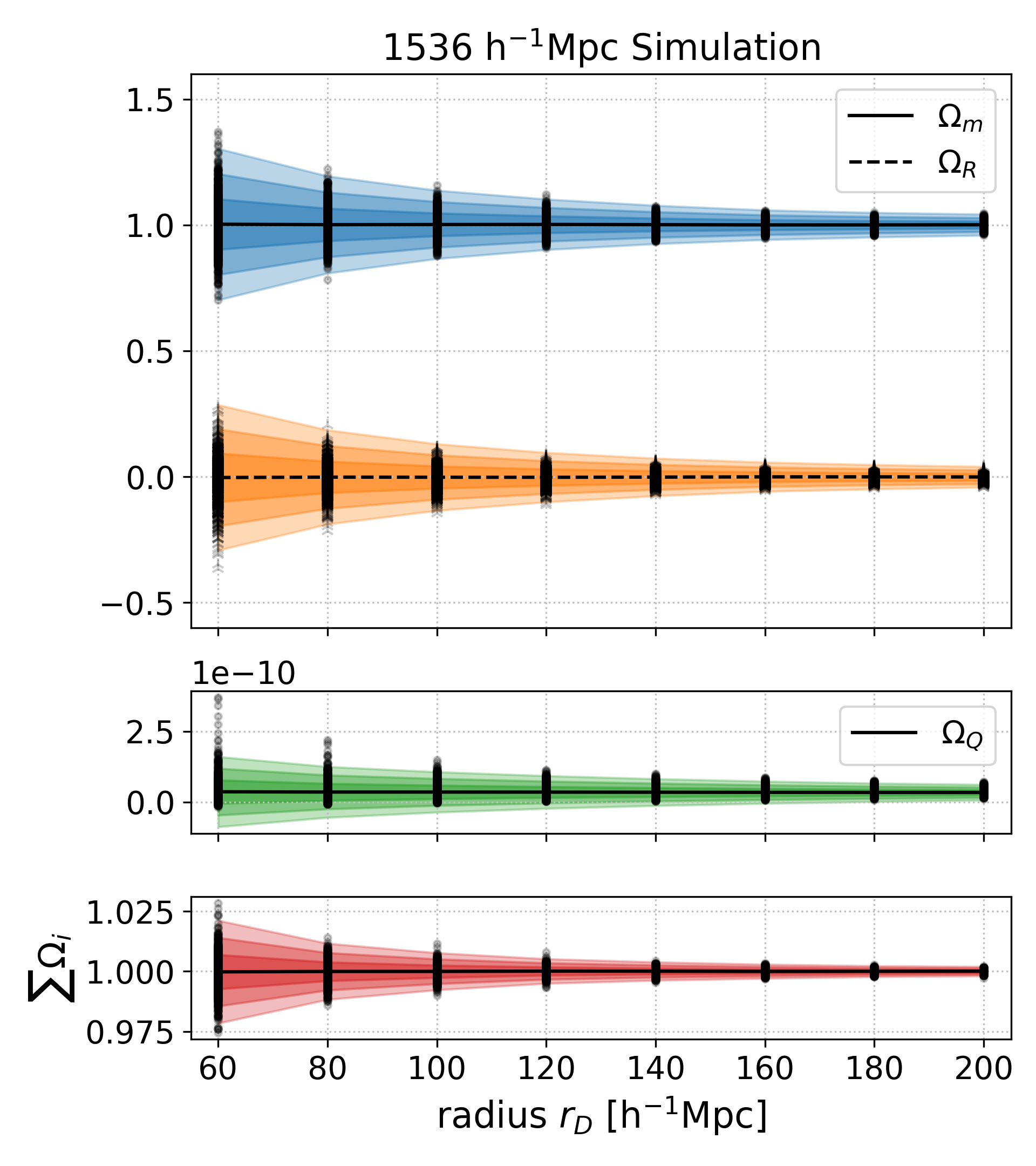}%
	\includegraphics[width=0.33\linewidth]{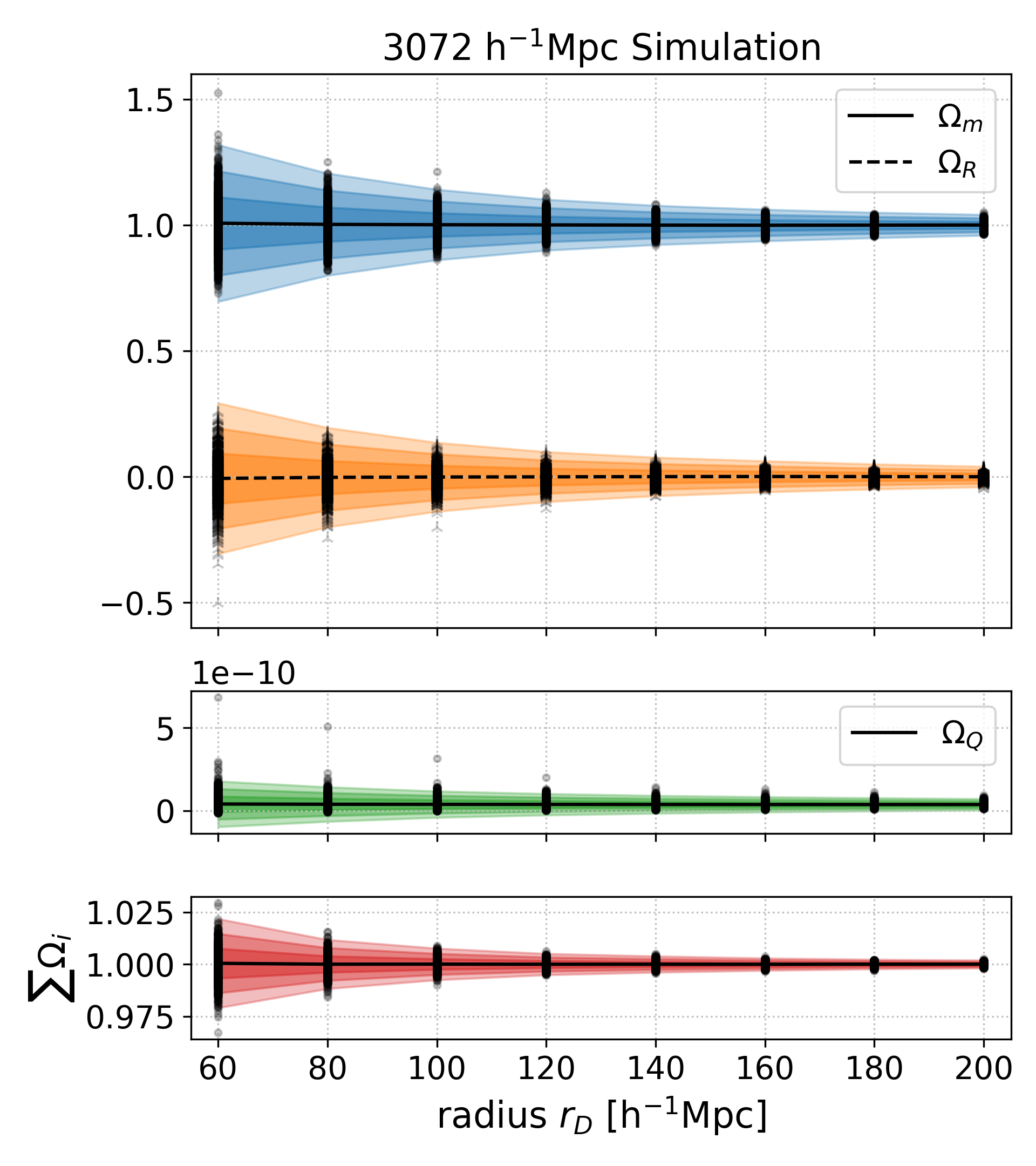}
	\caption{Average density parameters at the final simulation time point in 1000 randomly distributed sub-volumes of radius $r_\calD$ for the three different simulations considered in this paper. For each radius we plot the individual values, their mean and the 1,2 and 3$\sigma$ standard deviations. We also plot the sum of the individual density parameters as a measure of the average constraint violation.}
	\label{Fig:Omegas}
\end{figure*}

If we introduce the Hubble function 
\begin{equation}
	H (\tau) \equiv \frac{\p_\tau a}{a}
\end{equation}
and the time-dependent density parameters 
\begin{align}
	\Omega_m(\tau) &= \frac{8\pi G_N}{3 H^2}\langle\hat\rho^{(n)}\rangle\;, \\ 
	\Omega_R(\tau)&=-\frac{\langle{}^{(3)}\hat R\rangle}{6 H^2}\;,\\ 
	\Omega_Q(\tau)&=-\frac{\hat Q}{6 H^2}\;,
\end{align}
we can rewrite \eqref{eq:avg_friedm1_v2} as 
\begin{equation}
	1 = \Omega_m + \Omega_R + \Omega_Q\;.
\end{equation}
We further define
\begin{equation}
    \Omega_L(\tau) = \frac{2 \hat L}{3H^2}\;,
\end{equation}
such that \eqref{eq:avg_friedm2_v2} can be written as 
\begin{equation}
    \frac{\p_\tau^2 a}{a}\frac{1}{H^2} = -4\pi G_N\frac{\langle\hat p^{(n)}\rangle}{H^2}-\frac{\Omega_m}{2}+\frac{\Omega_Q}{2}+\frac{\Omega_L}{2},
\end{equation}
making it easy to compare the contribution of $\hat L$ to the other terms. For later convenience we also introduce the two variables 
\begin{equation}
    \mathcal{C}_q \equiv -\langle\hat\Gamma\hat\nabla_\alpha (\hat\Gamma^2 q^{(n)\alpha})\rangle\;,
\end{equation}
\begin{equation}
    \mathcal{C} \equiv \langle\hat\theta\rangle\langle\hat p\rangle-\langle\hat\theta\hat p^{(n)}\rangle  + \langle\rho^{(n)}\p_\tau\hat\Gamma^2\rangle +\mathcal{C}_q-\langle\hat\Gamma^3\sigma_{\alpha\beta}\pi^{(n)\alpha\beta}\rangle\;,
\end{equation}
such that the energy-conservation equation can be written as 
\begin{align}
	\p_\tau\langle\hat\rho^{(n)}\rangle+3\frac{\p_\tau a}{a}\left(\langle\hat\rho^{(n)}\rangle+\langle\hat p^{(n)}\rangle\right) = \mathcal{C}\;.
\end{align}
This equation can also be rewritten as 
\begin{align}
	\p_\tau\left(a^3\langle\hat\rho^{(n)}\rangle\right)= a^3 \left(\mathcal{C} -3\frac{\p_\tau a}{a}\langle\hat p^{(n)}\rangle\right) \;.
    \label{eq:avg_energy_consv_v3}
\end{align}

\subsection{Comment on the Constraint Violation}
Due to the finite grid size of the simulations, the Hamiltonian constraint \eqref{eq:Ham_const} is not perfectly fulfilled during the simulation. While the initial conditions are chosen such that \eqref{eq:Ham_const} is almost perfectly fulfilled on the initial hypersurface, it will be increasingly violated due to numerical inaccuracy on the later hypersurfaces. This problem will be worst in those areas, where highly non-linear structures build up, since the grid size is at some point inevitably no longer sufficient to resolve these structures.

To quantify how much the constraint is violated we define 
\begin{equation}
    \mathcal{H}={}^{(3)}R+K^2-K_{ij}K^{ij}-16\pi G_N \rho^{(n)}\;,
\end{equation}
which is equivalent to \eqref{eq:Ham_const} if $\mathcal{H}=0$. Non-zero values of $\mathcal{H}$ therefore quantify the constraint violation. We also define 
\begin{equation}
    |\mathcal{H}|=\sqrt{({}^{(3)}R)^2+K^4+(K_{ij}K^{ij})^2+(16\pi G_N \rho^{(n)})^2}
\end{equation}
and then plot the relative constraint violation $\mathcal{H}/|\mathcal{H}|$ in Fig.~\ref{Fig:constraint_violation}. We plot a single 2d-slice through each of the three simulation boxes and also plot the averaged $\mathcal{H}/|\mathcal{H}|$ in $8\times 8\times 8$ sub-boxes. While the constraint violation in individual boxes can be quite large, we can see that if averaged on multiple grid cells, the constraint violation is much reduced. To ensure that the average constraint is not violated too strongly we later only consider averages in spheres with minimal radius corresponding to 5 grid cells, which leads to spheres of roughly the same volume as the $8\times 8\times 8$ cubes. 

\begin{figure*}[]
	\centering
	\includegraphics[width=0.33\linewidth]{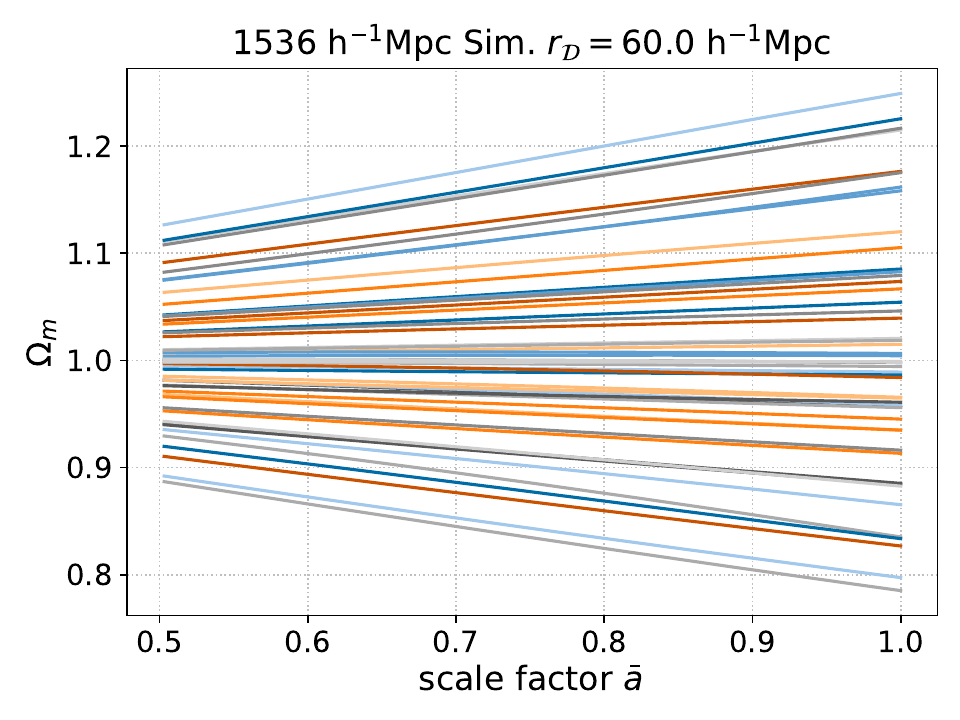}%
	\includegraphics[width=0.33\linewidth]{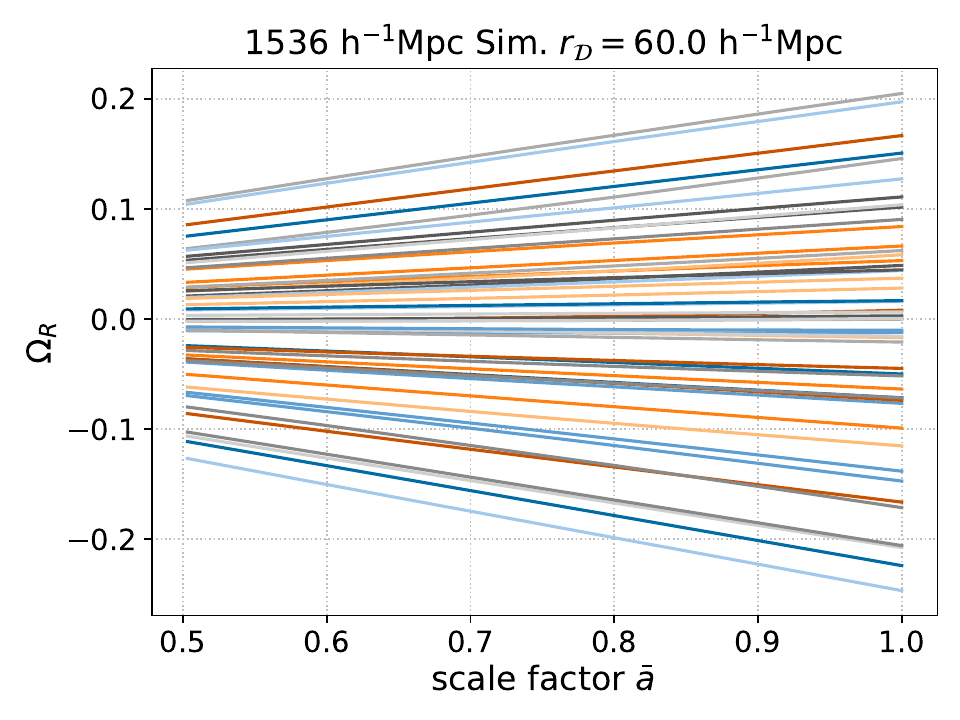}%
	\includegraphics[width=0.33\linewidth]{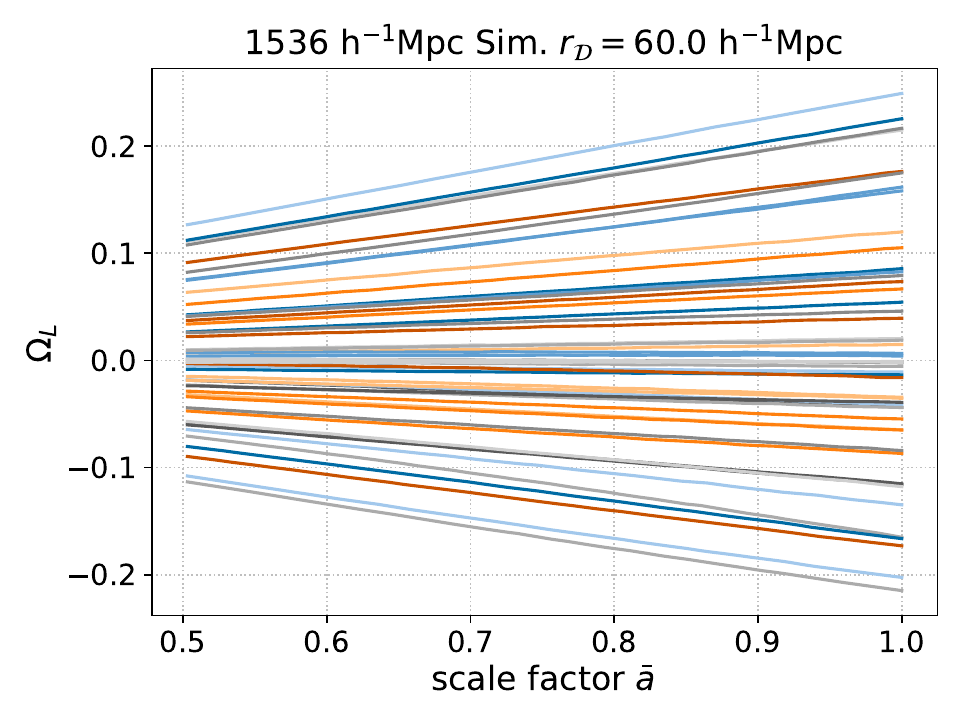}
	\includegraphics[width=0.33\linewidth]{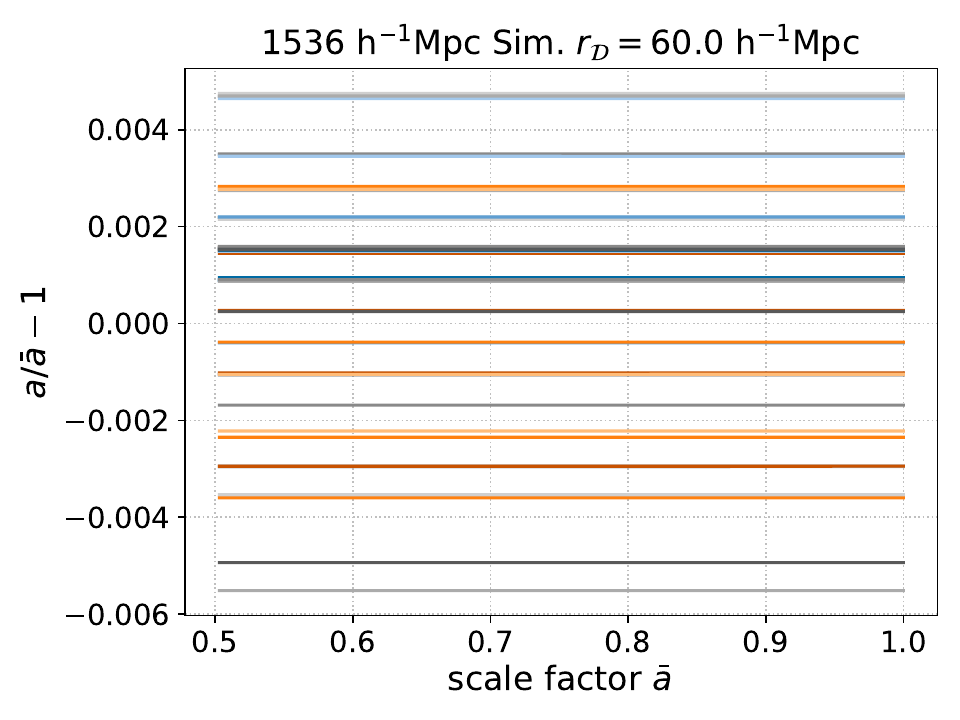}%
	\includegraphics[width=0.33\linewidth]{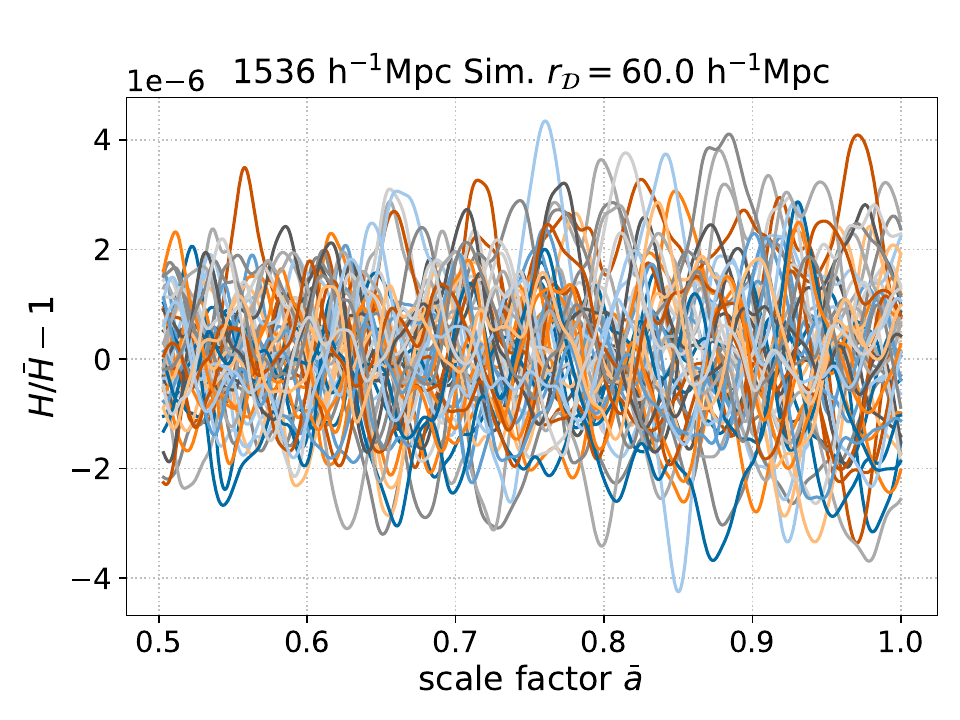}%
	\includegraphics[width=0.33\linewidth]{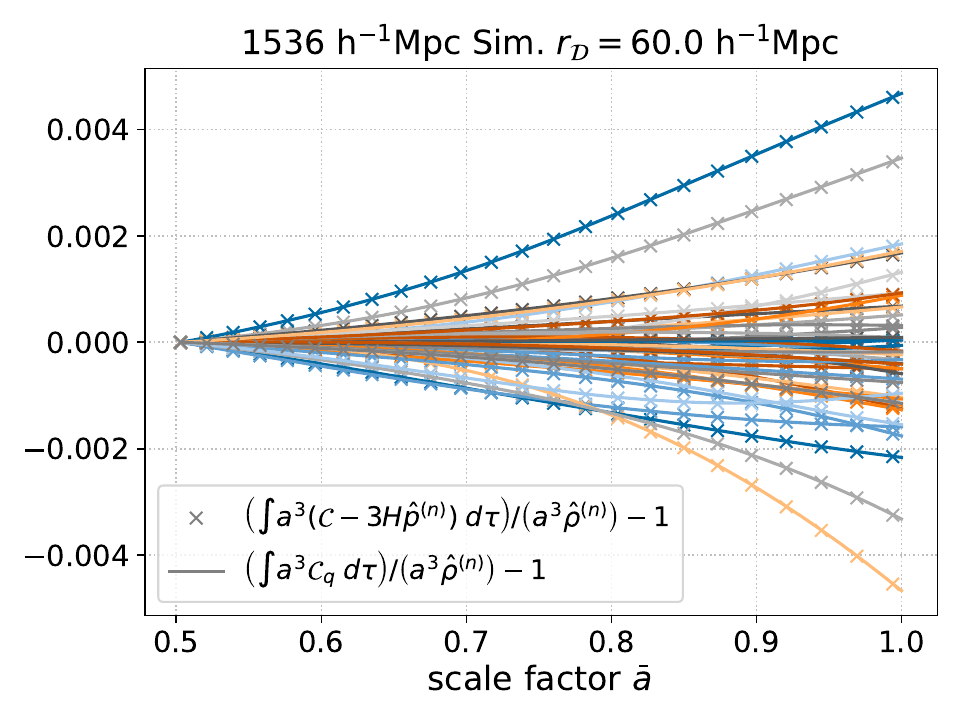}
	\includegraphics[width=0.33\linewidth]{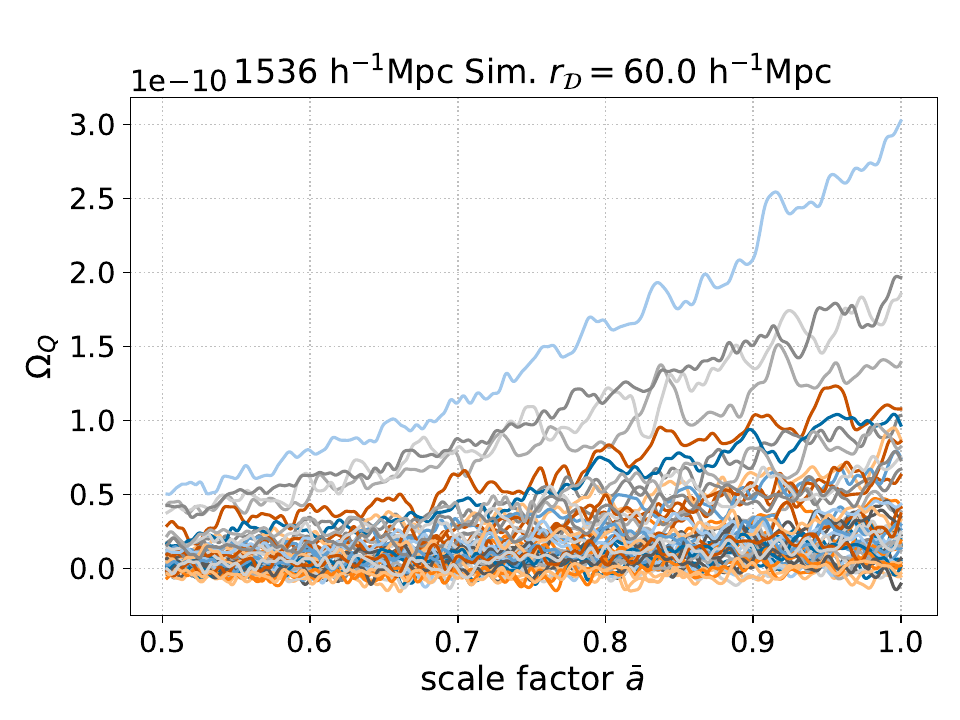}%
	\includegraphics[width=0.33\linewidth]{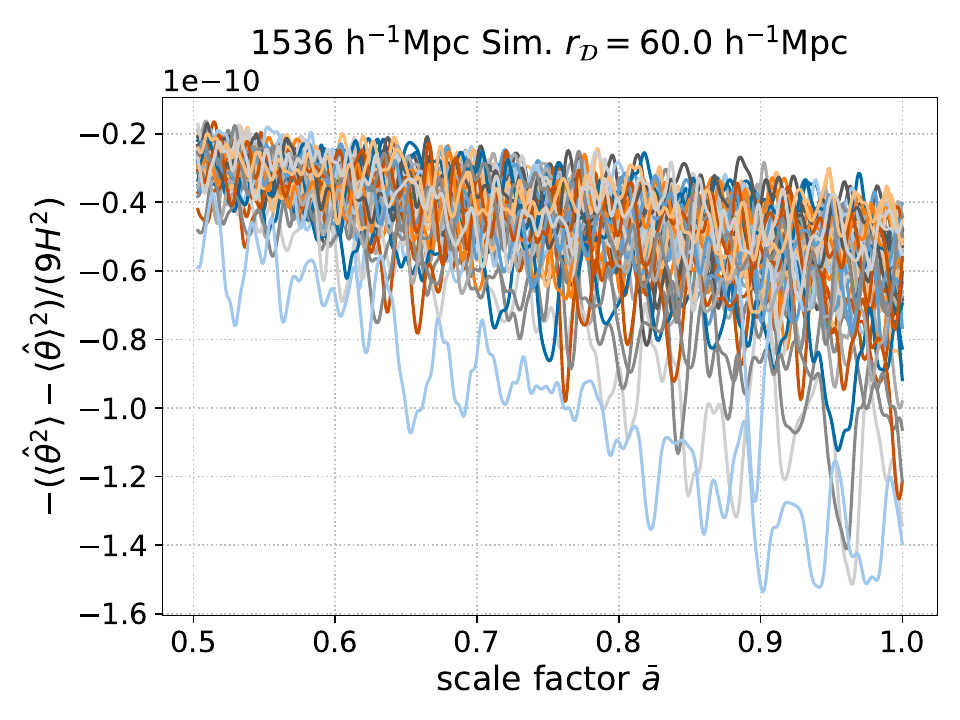}%
	\includegraphics[width=0.33\linewidth]{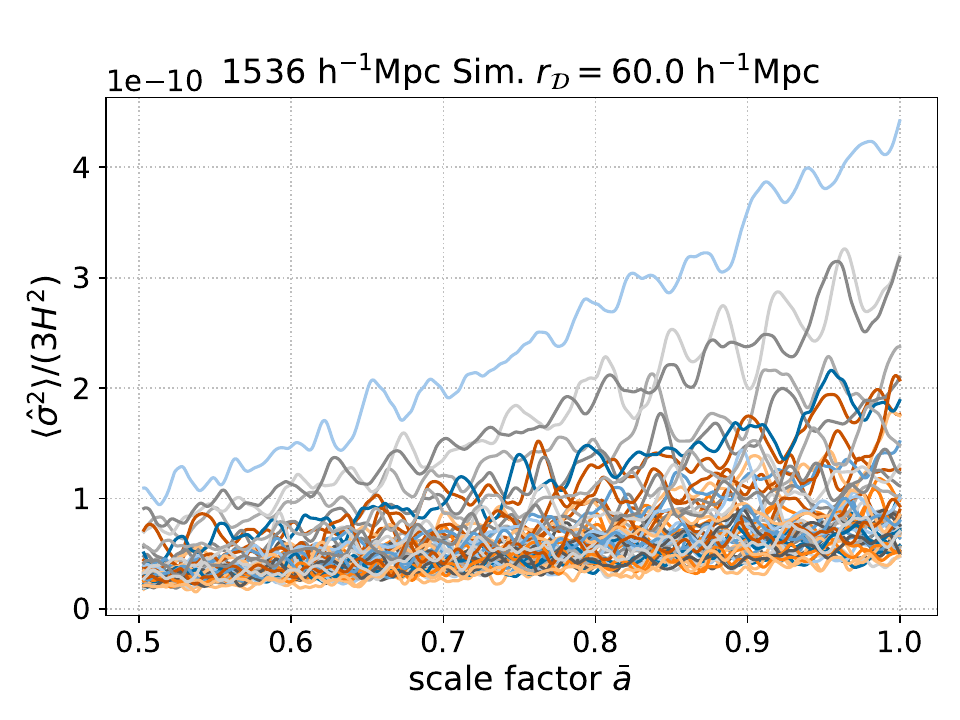}
    \caption{Different averaged quantities in 50 randomly selected spheres of radius $r_\calD=60\;\mathrm{h}^{-1}\mathrm{Mpc}$ in the simulation with side length $1536\;\mathrm{h}^{-1}\mathrm{Mpc}$ plotted versus the mean scale factor $\bar a$ of the whole box.}
    \label{Fig:time_evol_1536_60}
\end{figure*}

\begin{figure*}[]
	\centering
	\includegraphics[width=0.33\linewidth]{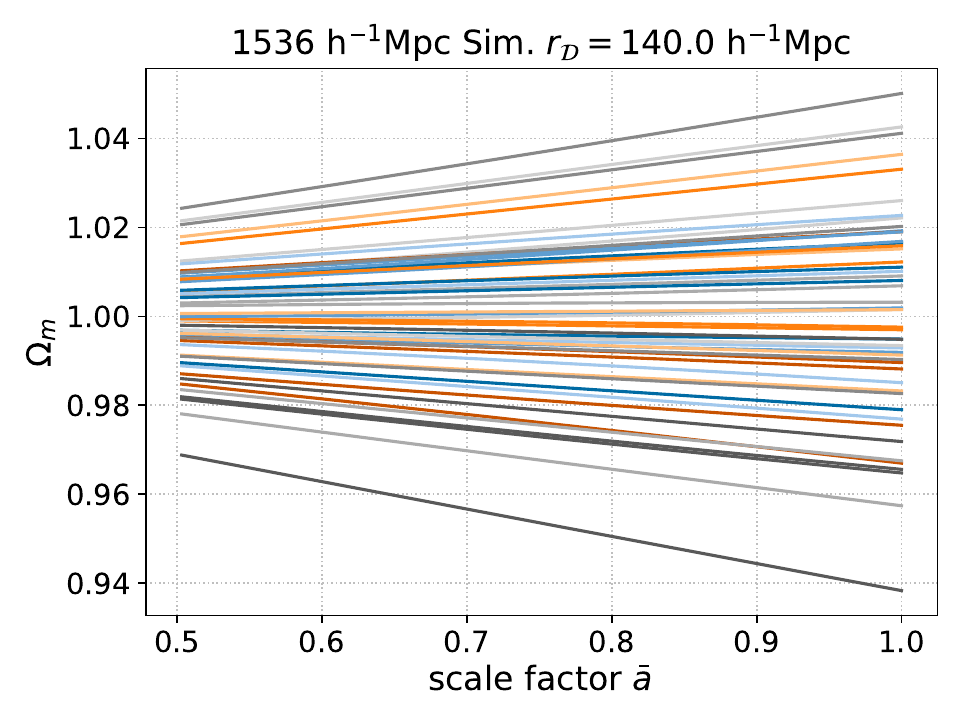}%
	\includegraphics[width=0.33\linewidth]{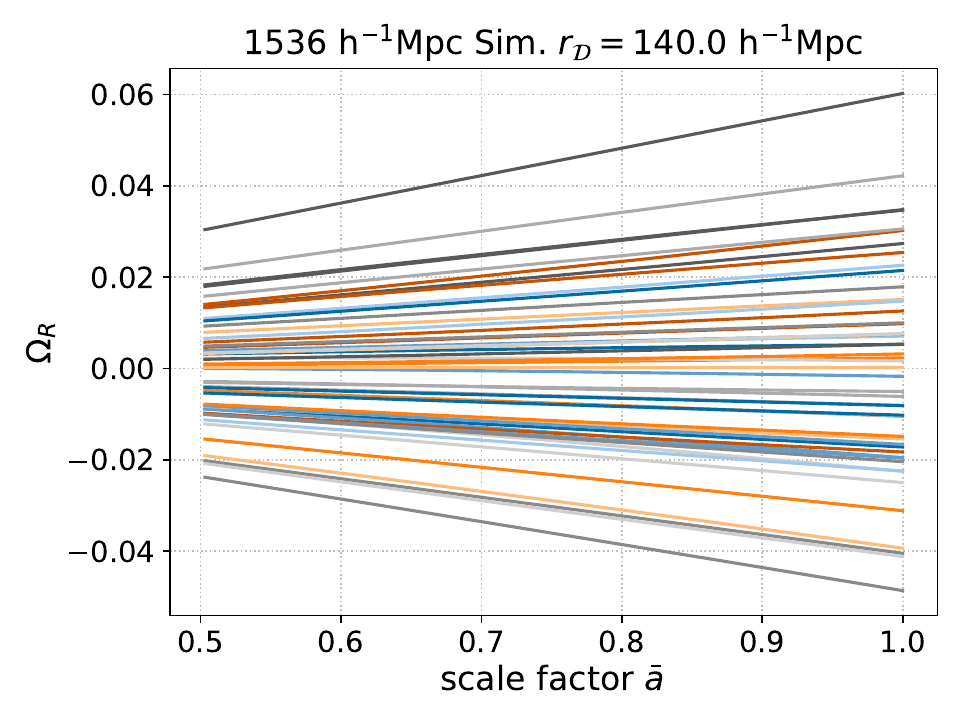}%
	\includegraphics[width=0.33\linewidth]{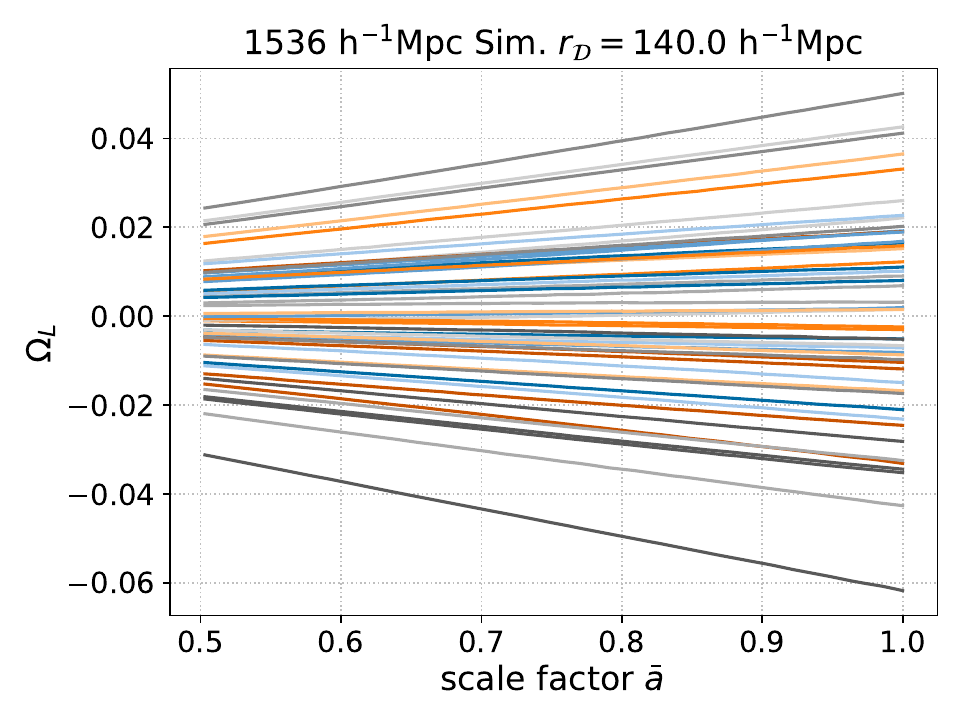}
	\includegraphics[width=0.33\linewidth]{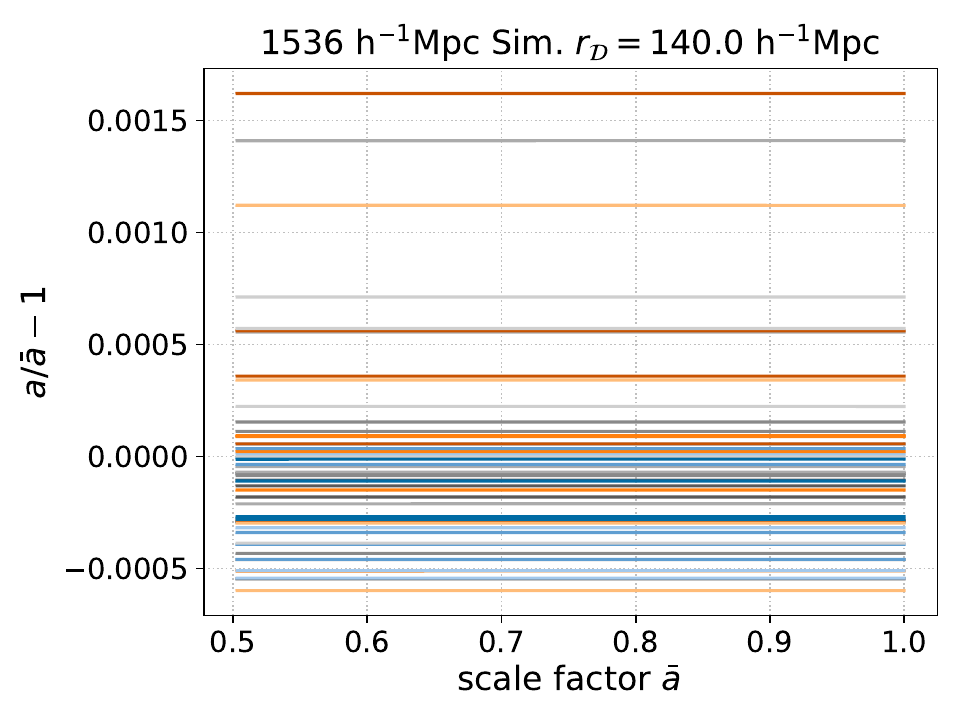}%
	\includegraphics[width=0.33\linewidth]{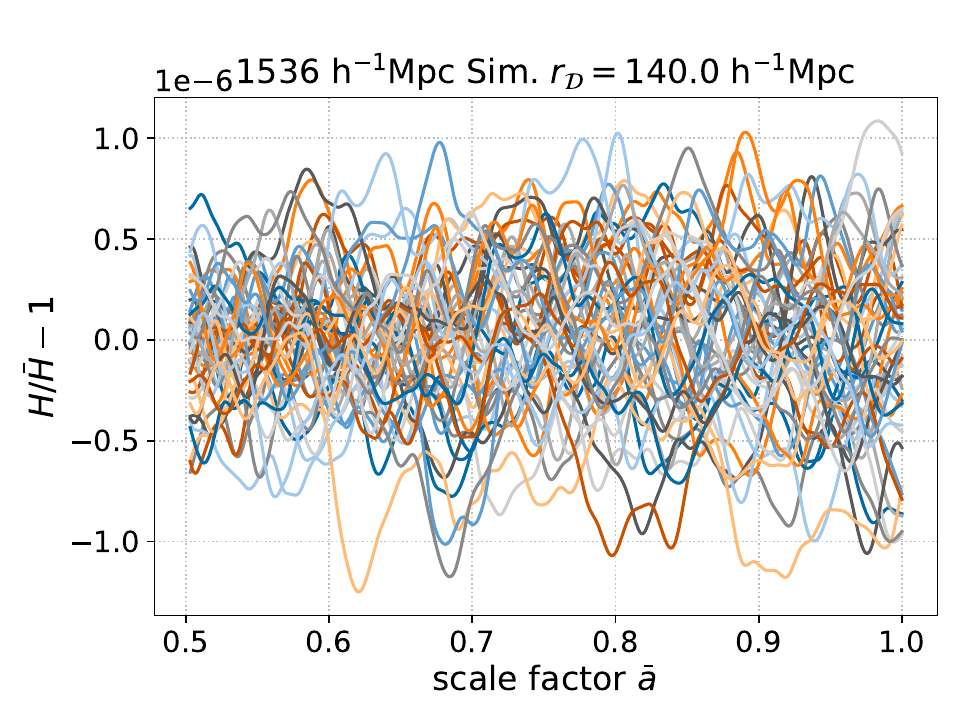}%
	\includegraphics[width=0.33\linewidth]{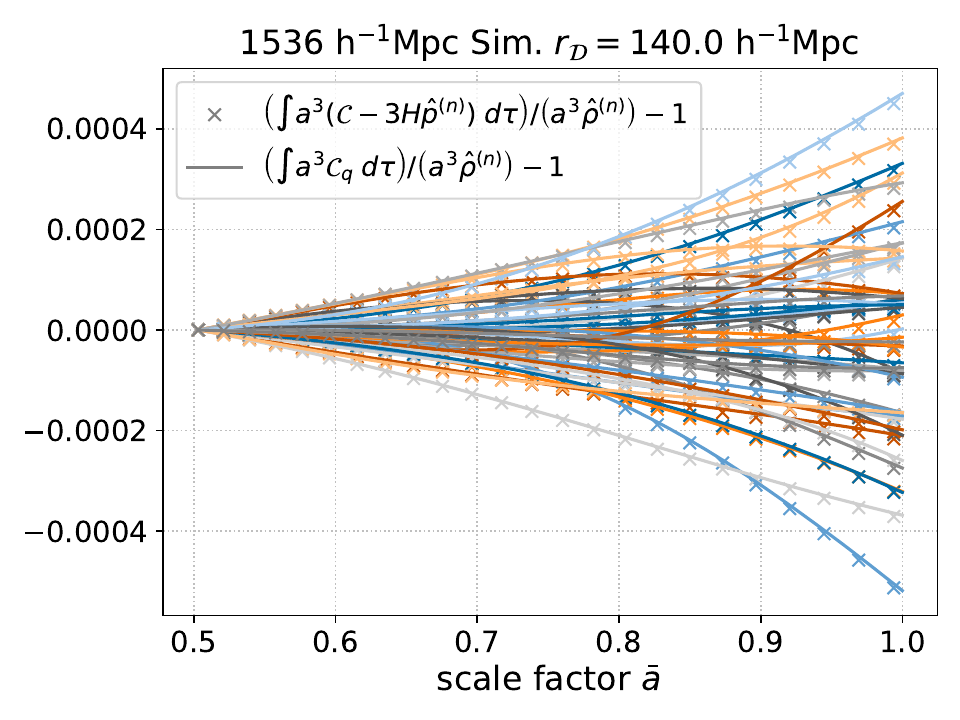}
	\includegraphics[width=0.33\linewidth]{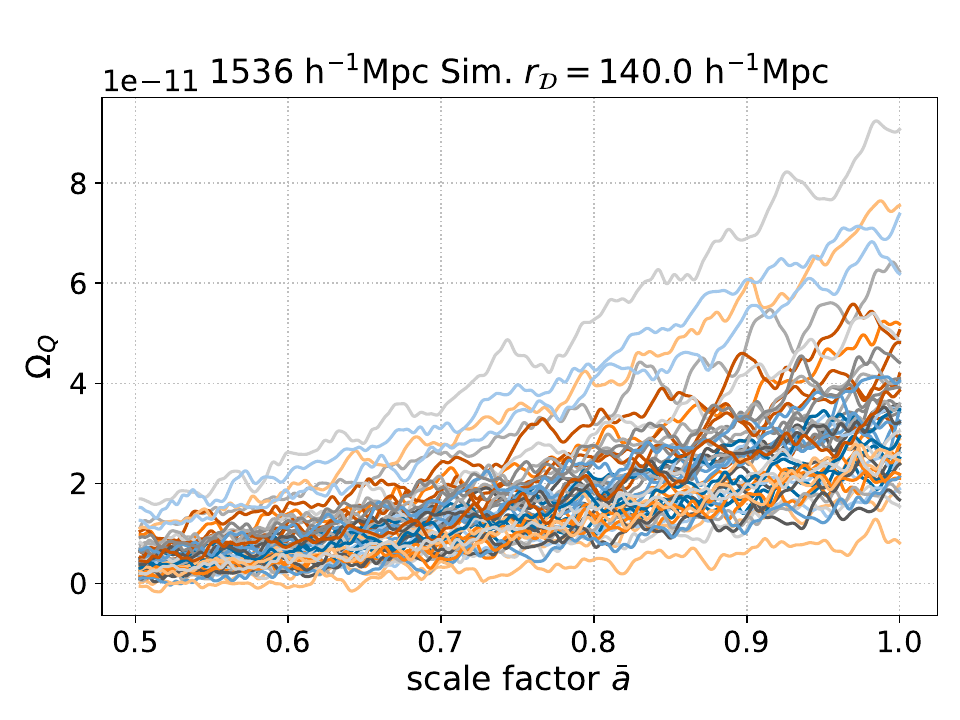}%
	\includegraphics[width=0.33\linewidth]{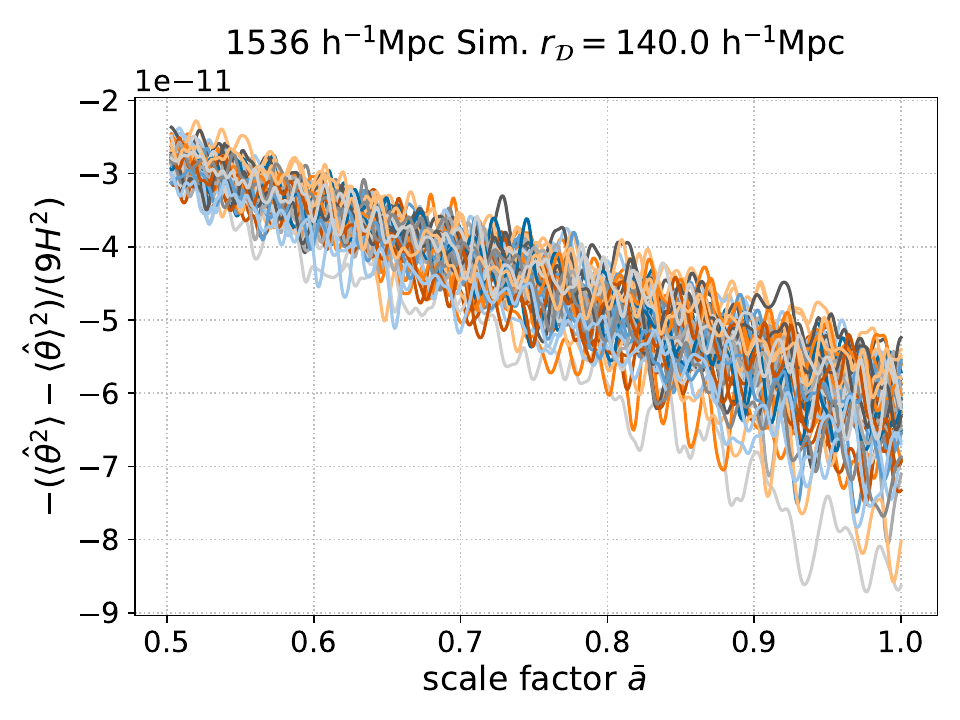}%
	\includegraphics[width=0.33\linewidth]{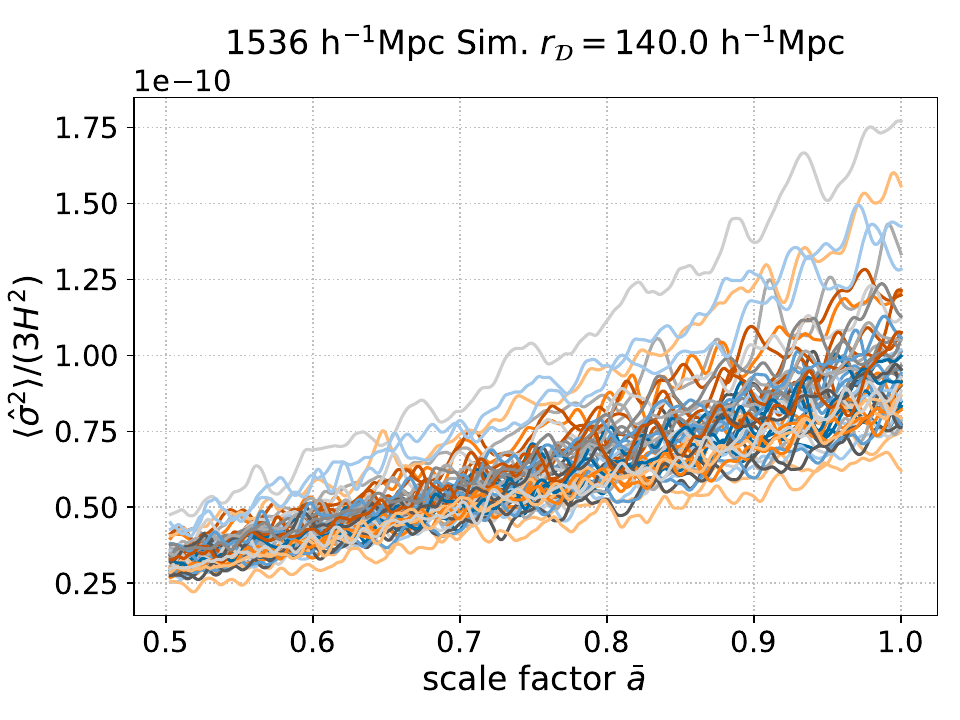}
    \caption{Different averaged quantities in 50 randomly selected spheres of radius $r_\calD=140\;\mathrm{h}^{-1}\mathrm{Mpc}$ in the simulation with side length $1536\;\mathrm{h}^{-1}\mathrm{Mpc}$ plotted versus the mean scale factor $\bar a$ of the whole box.}
    \label{Fig:time_evol_1536_140}
\end{figure*}
\begin{figure}
    \centering
    \includegraphics[width=0.8\linewidth]{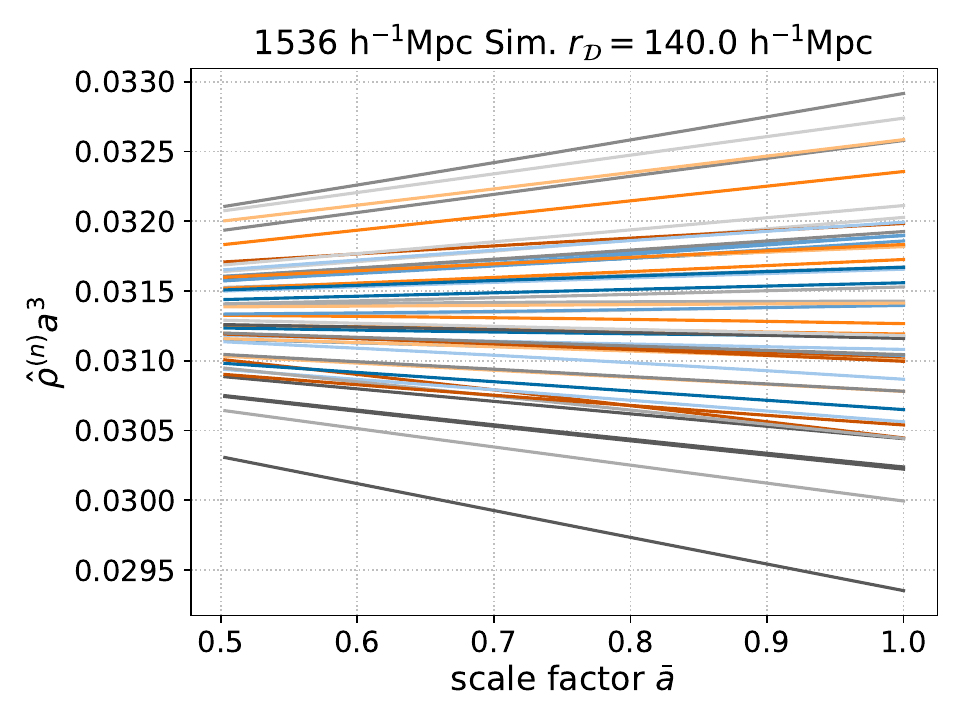}
    \includegraphics[width=0.8\linewidth]{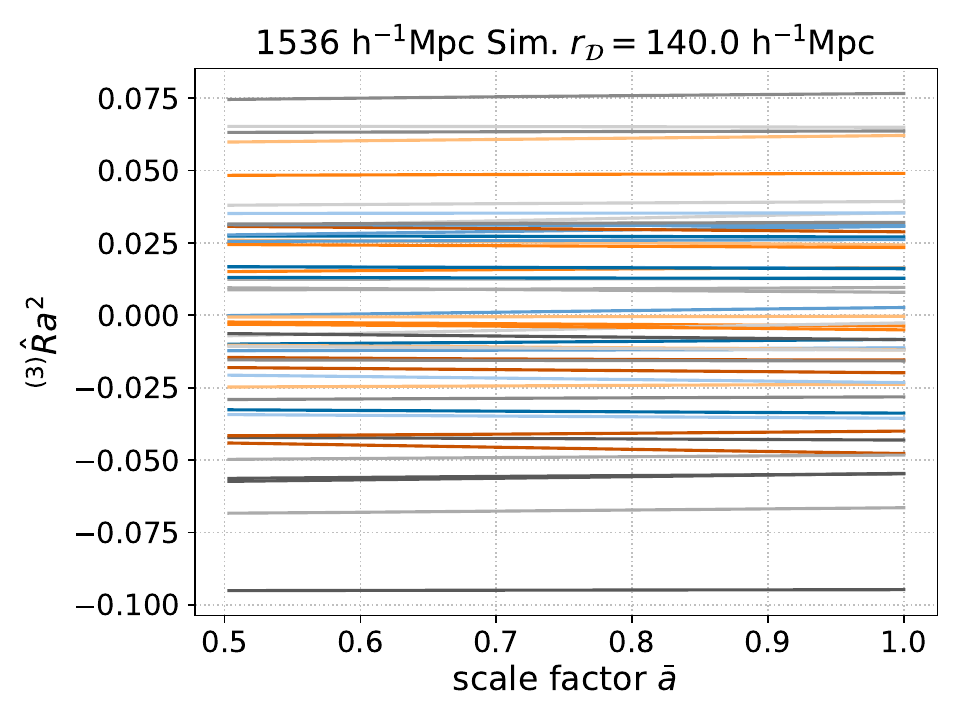}
    \caption{Average density and curvature in 50 randomly selected spheres of radius $r_\calD=140\;\mathrm{h}^{-1}\mathrm{Mpc}$ in the simulation with side length $1536\;\mathrm{h}^{-1}\mathrm{Mpc}$ plotted versus the mean scale factor $\bar a$ of the whole box, scaled to illustrate their deviation from standard FLRW evolution.}
    \label{fig:rho_curvature}
\end{figure}

\section{Results}
\label{sec:4}
We study three different simulations. Two simulation have $128^3$ grid points each and box sizes $512\;\mathrm{h}^{-1}\mathrm{Mpc}$ and $1536\;\mathrm{h}^{-1}\mathrm{Mpc}$, respectively. The remaining simulation has $256^3$ grid points and box size $3072\;\mathrm{h}^{-1}\mathrm{Mpc}$. For each simulation we calculate average quantities in 1000 randomly chosen spheres for a range of different radii. The two larger simulations have the same physical resolution, meaning the $1536\;\mathrm{h}^{-1}\mathrm{Mpc}$ is statistically equivalent to a 1/8 sub-box of the $3072\;\mathrm{h}^{-1}\mathrm{Mpc}$ simulation. We show results from both simulations, since we have less overlap of individual averaging spheres in the larger simulation and will also have spheres further away from the boundary. Providing at least some measure of the influence of the periodic boundary conditions on the result.
\newline

We start our analysis by looking at the averaged equivalent of the first Friedmann equation in the form \eqref{eq:avg_friedm1_v2}. In Fig.~\ref{Fig:Omegas} we show the three density parameters $\Omega_m$, $\Omega_R$ and $\Omega_Q$ in each of the three simulations, for all spheres and radii together with their mean and  $1\sigma$, $2\sigma$ and $3\sigma$ standard deviations at the final time point of the simulation. We find that the backreaction is negligible in all the simulations, at all averaging scales. The quantity $\Omega_Q$ is at most of the order $10^{-9}$ for spheres with radius $r_\calD = 20\;\mathrm{h}^{-1}\mathrm{Mpc}$ in the smallest simulation box. However, we see that in individual spheres, the average matter and curvature density parameters can deviate significantly from one and zero, respectively, as we would expect in the late universe where structure has built up in over-dense regions and voids have formed. We note that $\Omega_R$ varies up to $10\%$ in spheres as large as $r_\calD \sim 200\;\mathrm{Mpc}$, close to (if not above) the suspected homogeneity scale. The average over all density parameters in the different spheres agrees with the average of the entire box. The only exception to this is for the largest radii in the smallest box. Here, the average $\Omega_m$ is slightly larger than one. This is caused by the fact that \texttt{mescaline} picks random origins for the spheres such that the spheres never touch the edge of the simulation box. If we choose a radius that comes close to the box size, all spheres are therefore centered around the center of the whole simulation volume, which happens to be slightly over-dense in our case. This could be fixed in the future by making use of the periodic boundary conditions of the simulation, but since this is not our main focus and not a problem for the larger simulations we use, we leave this to future work. We also show the sum of the different density parameters in Fig.~\ref{Fig:Omegas}, corresponding to the average constraint violation in the individual spheres. For all spheres larger than $r_\calD = 50\;\mathrm{h}^{-1}\mathrm{Mpc}$ it is smaller than a few percent and always smaller then the deviation of $\Omega_m$ ($\Omega_R$) from one (zero). We therefore conclude that it is unlikely that errors due to constraint violation dominate our results. The constraint violation is, however, many orders of magnitude larger than $\Omega_Q$, so results regarding $\Omega_Q$ should be taken with some caution, but we will comment on this again further below. The results from all three simulation boxes are of the same order of magnitude, reassuring us that the different resolutions and box sizes do not influence the result. 
\newline

We now move on to consider the time evolution of the different averaged quantities for selected radii. The $1536\;\mathrm{h}^{-1}\mathrm{Mpc}$ and $3072\;\mathrm{h}^{-1}\mathrm{Mpc}$ both have significantly smaller constraint violation in the individual grid-cells than the smaller simulation and in addition, the two bigger simulations have outputs for a larger range of scale factors, i.e. a larger time span of the universes evolution. Since the computational cost of calculating averages with \texttt{mescaline} scales with the number of grid points in the entire simulation box and the results for all box sizes are in good qualitative agreement we consider only the $1536\;\mathrm{h}^{-1}\mathrm{Mpc}$ box in the following. We consider the two averaging radii $r_\calD= 60, 140\;\mathrm{h}^{-1}\mathrm{Mpc}$ which represent the (nearly) smallest and largest radii we have studied, but we note that the results for the remaining radii are qualitatively similar. We show evolution of the density parameters in Figs.~\ref{Fig:time_evol_1536_60} and \ref{Fig:time_evol_1536_140}. The results for the two different averaging radii agree qualitatively and we therefore describe them together in the following. Quantitatively the averages in the smaller spheres in Fig.~\ref{Fig:time_evol_1536_60} display slightly more backreaction and matter in- and outflow in the spheres. We show these results for only 50 (randomly chosen) of our original 1000 spheres in order to retain readability of the figures. Any gaps or uneven distribution seen in the figure is simply due to this random selection and not present when plotting results for all 1000 spheres.

In the top row of each figure, we plot the three density parameters $\Omega_m$, $\Omega_R$ and $\Omega_L$. We can see that $\Omega_m$ changes over time, such that over-dense regions become even denser and under-dense regions loose more matter. As demonstrated explicitly in Fig.~\ref{fig:rho_curvature}, the average matter density and curvature do not evolve exactly as in FLRW spacetimes, i.e. as proportional to $a^{-3}$ and $a^{-2}$, respectively. However, we demonstrate below that this behavior is entirely due to matter in- and outflow of our averaging spheres.

From Figs.~\ref{Fig:time_evol_1536_60} and \ref{Fig:time_evol_1536_140} we see that over-dense regions (as expected) are positively curved (negative $\Omega_R$) and under-dense regions negatively curved (positive $\Omega_R$), with $\Omega_R$ changing in an equal but opposite manner to $\Omega_m$ (up to the constraint violation). $\Omega_L$ changes exactly such that there is no net change in the acceleration equation \eqref{eq:avg_friedm2_v2}, i.e. $\Omega_L$ cancels the non-constant part in $\Omega_m$. The acceleration of $a$ is therefore not affected by the in- and outflow and the second derivative of $a$ behaves as in a standard (near-)EdS universe. 

The first two plots in the second line of Figs.~\ref{Fig:time_evol_1536_60} and \ref{Fig:time_evol_1536_140} show the relative difference in the sub-volume averaged scale factor and Hubble function compared to the whole box average. We can see that there is a slight difference in the scale factors in the individual sub-volumes, but the relative difference is constant over time and the Hubble function remains unchanged. The relative difference in the Hubble function is of the order $10^{-6}$ and oscillates wildly. We interpret the oscillations to mean that we have reached the numerical accuracy of the simulations, caused by the finite grid size of the simulations (we discuss further evidence of this at the end of the section). Note that this means that the actual relative difference could be even smaller and that the deviations we find in the Hubble parameter should be considered upper limits.

In order to confirm that that the matter-density times $a^3$ and curvature times $a^2$ in the individual spheres are changing due solely to matter in- and outflow of the spheres, we consider the energy-momentum conservation equation \eqref{eq:avg_energy_consv_v2} in the form \eqref{eq:avg_energy_consv_v3}. Integrating the right hand side of the equation equals $a^3\hat\rho^{(n)}$ which is constant if the density changes only due to the expansion of space. We integrate the entire right hand side and also consider the integral over only the term $\mathcal{C}_q$ appearing in $\mathcal{C}$. For both these integrals, we plot the relative difference to $a^3\hat\rho^{(n)}$. These plots are the last plots in the second line in Figs.~\ref{Fig:time_evol_1536_60},\ref{Fig:time_evol_1536_140}. The results obtained from using the two different integrals agree to high precision and we cannot distinguish the two different types of curves in the figure. The relative difference to $a^3\hat \rho^{(n)}$ is of $<1\%$, demonstrating that the change in the density which is not due to the expansion of space, is due to $\mathcal{C}_q$. The relative difference is not exactly zero due to the constraint violation already discussed. $\mathcal{C}_q$ is proportional to a gradient in the average re-scaled energy flux $\hat\Gamma^2q^{(n]}_\alpha$, and therefore matter flowing in and out of the spheres. Equation \eqref{eq:LRQC_rel} in combination with \eqref{eq:avg_energy_consv_v2} shows that $L, a^6 Q, a^2R$ are sourced by $\mathcal{C}$, where in our case $\mathcal{C}_q$ is the dominant part.

In the last row of Figs.~\ref{Fig:time_evol_1536_60},\ref{Fig:time_evol_1536_140} we plot $\Omega_Q$.  We also plot the two terms constituting $\Omega_Q$ individually in order to demonstrate that the two terms are individually small (as opposed to simply cancelling). The dominant contribution comes from the shear. Numerically, $\Omega_Q$ as well as its two constituent terms increase slightly over time and in addition, $\Omega_Q$ is slightly larger in smaller spheres. However, all plots of these quantities oscillate similar to the relative change in $H$, although not quite as much. Since $\Omega_Q$ is proportional to $\theta^2$ and thereby $H^2$, the results are actually slightly above the suspected noise level estimated from the relative deviations in $H$. We nonetheless assess that it is uncertain to what extent we can trust the trends of increasing backreaction at smaller scales and later times. The main message of the plots of $\Omega_Q$ are therefore that this quantity is negligibly small.

\begin{figure*}[]
	\centering
	\includegraphics[width=0.33\linewidth]{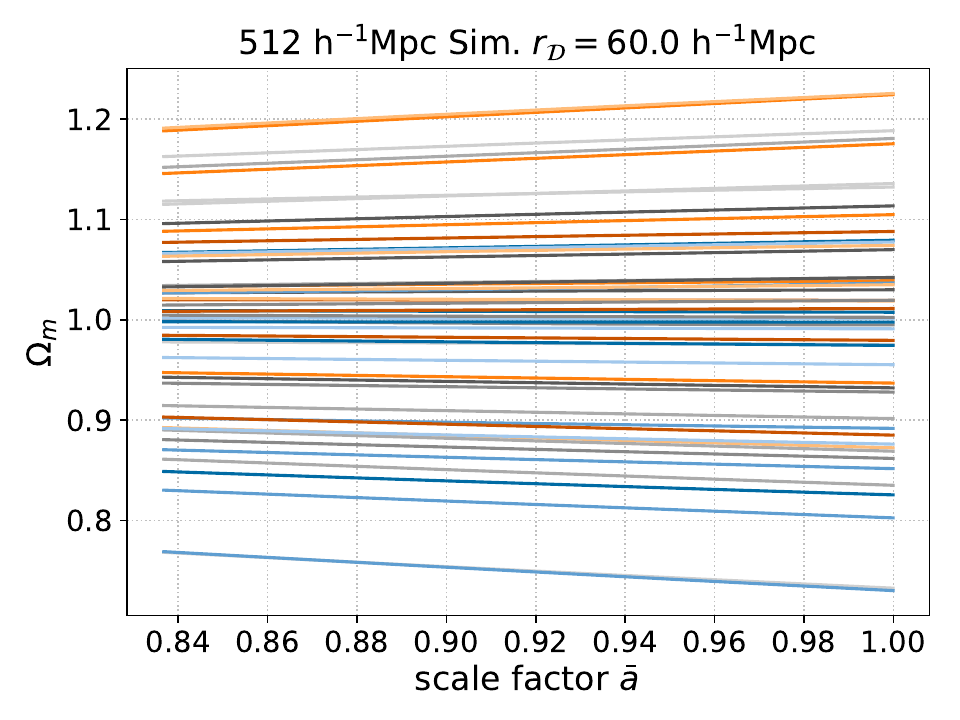}%
	\includegraphics[width=0.33\linewidth]{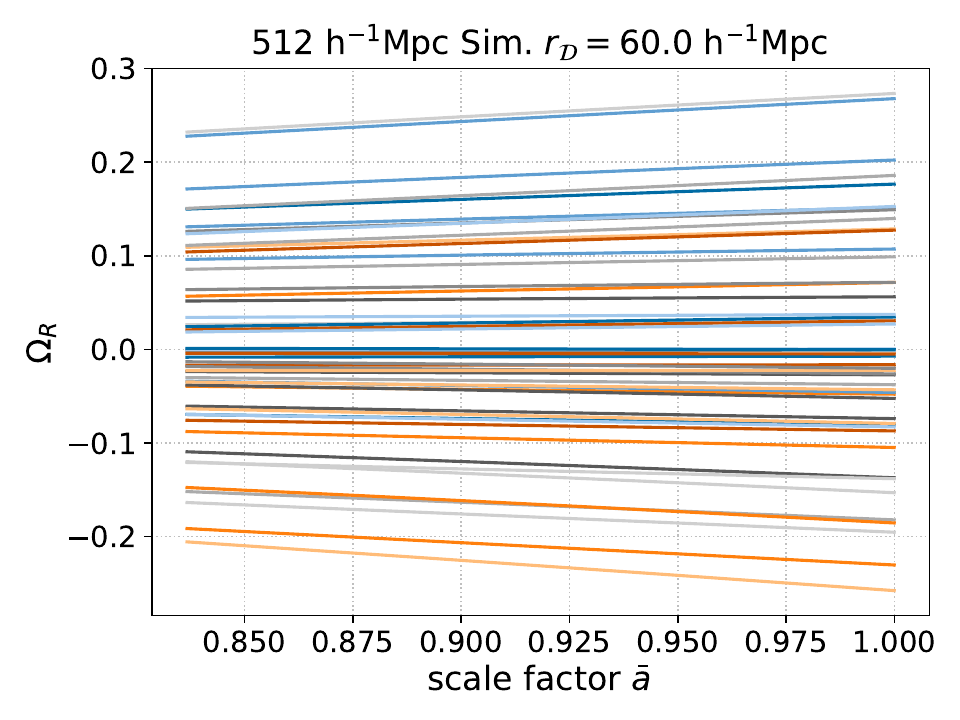}%
	\includegraphics[width=0.33\linewidth]{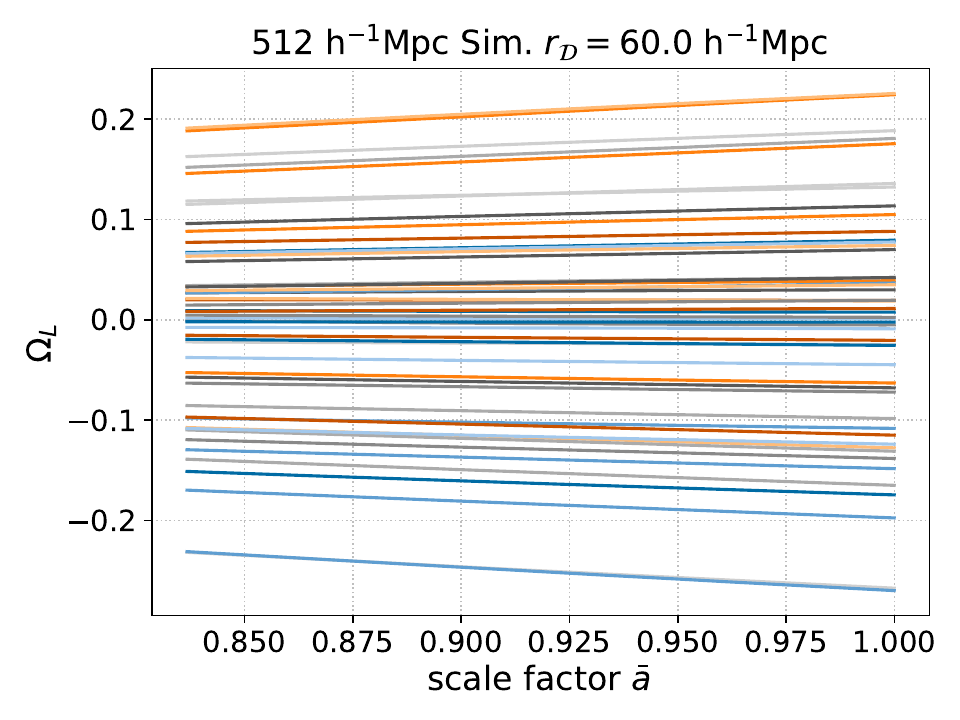}
	\includegraphics[width=0.33\linewidth]{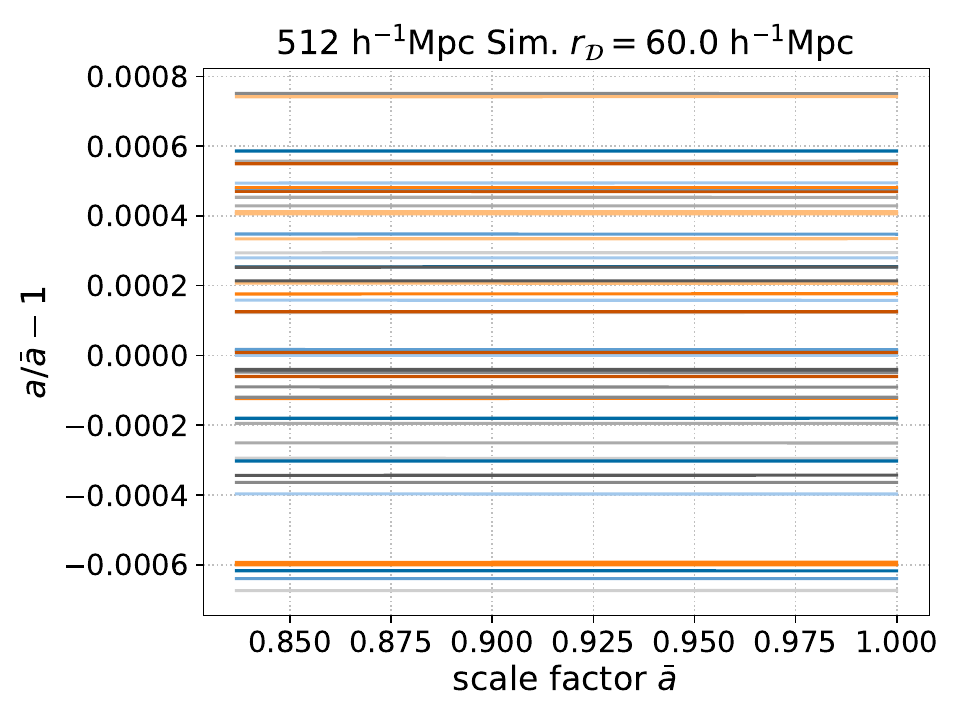}%
	\includegraphics[width=0.33\linewidth]{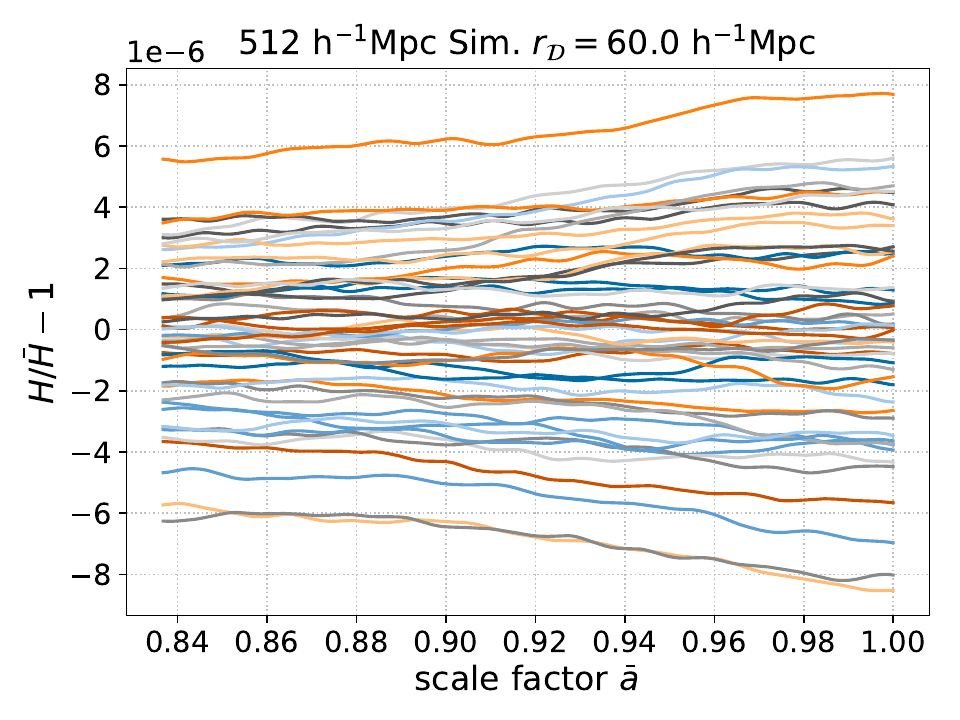}%
	\includegraphics[width=0.33\linewidth]{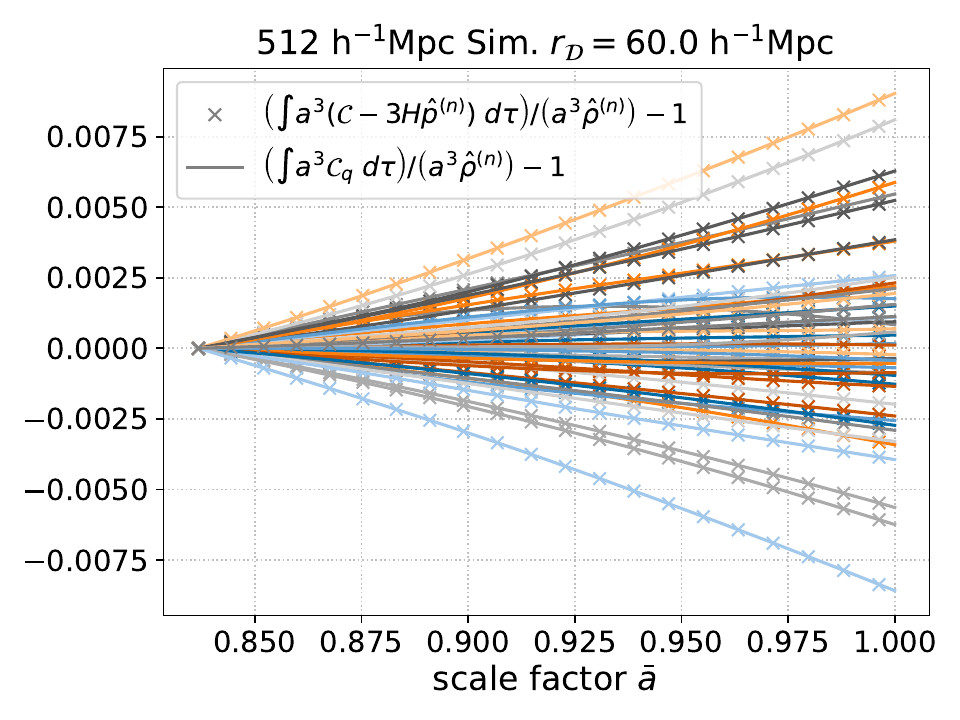}
	\includegraphics[width=0.33\linewidth]{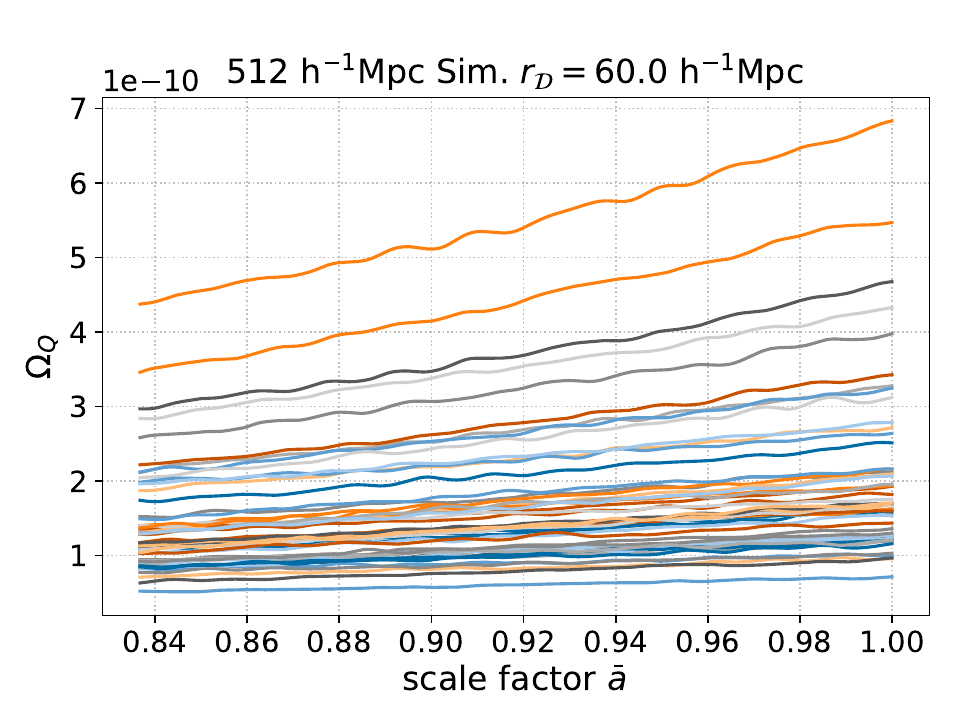}%
	\includegraphics[width=0.33\linewidth]{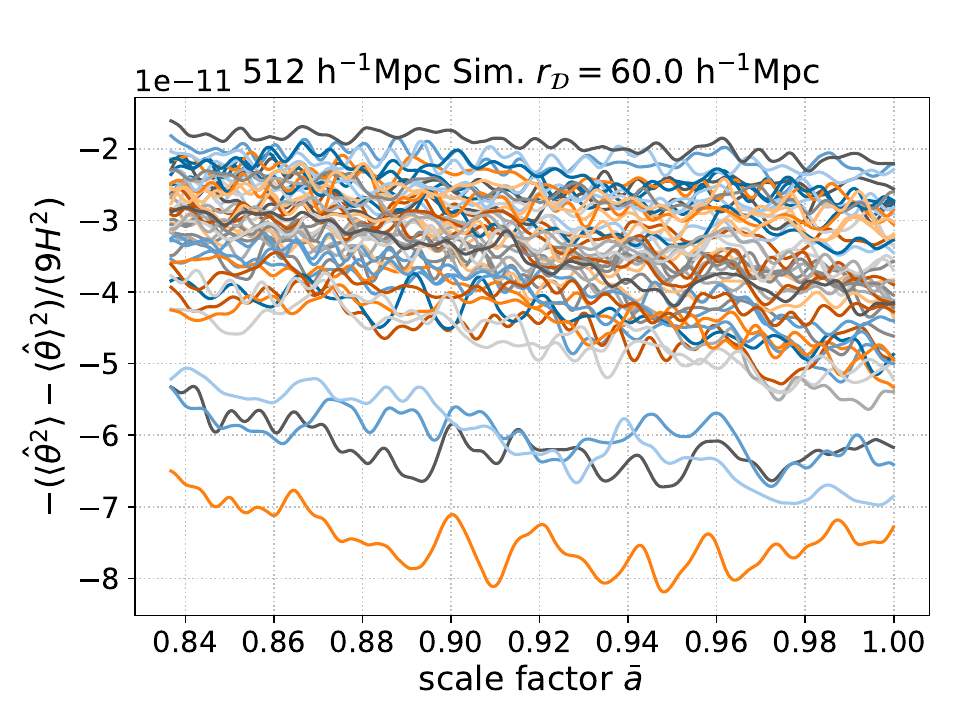}%
	\includegraphics[width=0.33\linewidth]{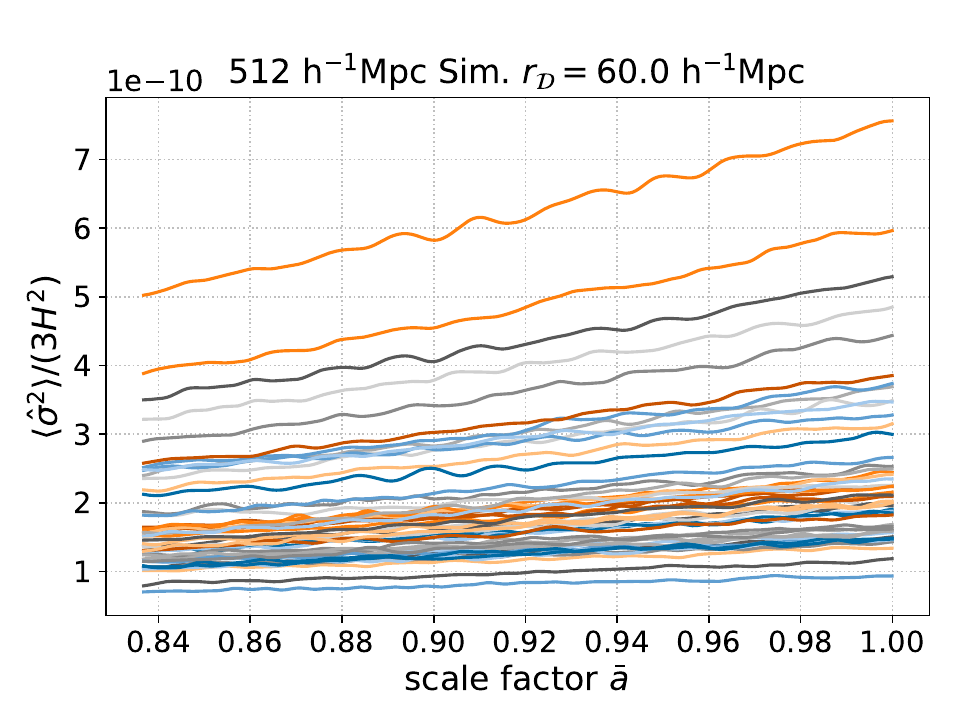}
    \caption{Different averaged quantities in 50 randomly selected spheres of radius $r_\calD=60\;\mathrm{h}^{-1}\mathrm{Mpc}$ in the simulation with side length $512\;\mathrm{h}^{-1}$Mpc plotted versus the mean scale factor $\bar a$ of the whole box. }
    \label{Fig:time_evol_512_60}
\end{figure*}

\begin{figure}[]
	\centering
	\includegraphics[width=0.8\linewidth]{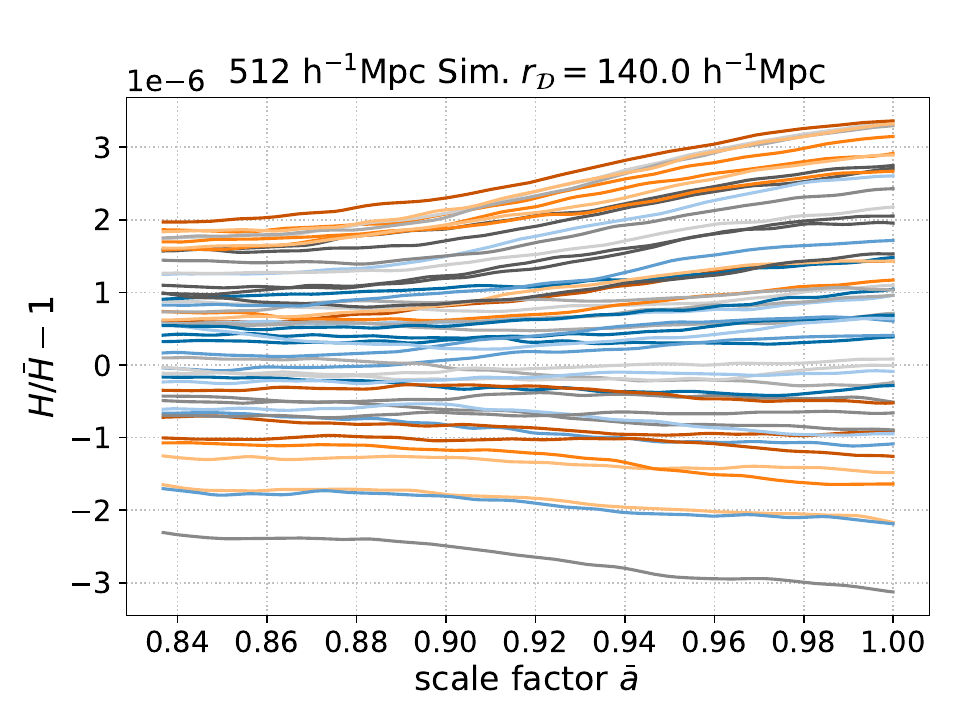}
    \caption{Relative difference in the Hubble rate $H$ compared to the whole box average $\bar H$ in 50 randomly selected spheres of radius $r_\calD=140\;\mathrm{h}^{-1}\mathrm{Mpc}$ in the simulation with side length $512\;\mathrm{h}^{-1}$Mpc plotted versus the mean scale factor $\bar a$ of the whole box. }
    \label{Fig:time_evol_512_140}
\end{figure}

We also show similar plots as those just discussed for one averaging radius in the $512\;\mathrm{h}^{-1}\mathrm{Mpc}$ simulation box in Fig.~\ref{Fig:time_evol_512_60}. We choose $r_\calD = 60\;\mathrm{h}^{-1}\mathrm{Mpc}$ to have a direct comparison to Fig.~\ref{Fig:time_evol_1536_140}. The most notable difference is that now the relative difference in the Hubble function $H$ does not oscillate as much and shows a slight trend. A possible explanation for the ``missing'' fluctuations could be the higher resolution of the smaller simulation box, supporting our earlier expectation that the fluctuations are caused by the finite grid resolution of the simulations. To further support this, we also show the relative fluctuation in the Hubble rate for the $512\;\mathrm{h}^{-1}\mathrm{Mpc}$ simulation box and the radius $r_\calD = 140\;\mathrm{h}^{-1}\mathrm{Mpc}$ in Fig.~\ref{Fig:time_evol_512_140}. We see that the oscillations are now completely absent and we have a clear signal. $\Omega_Q$ and its constituents also oscillate significantly less. All other results are consistent between the different simulation boxes.

\subsection{Relation to other Works}
Spatial averages and backreaction have earlier been studied in ET simulations in \cite{Macpherson2019} and \cite{Williams2024}. Both of these papers consider different averaging formalisms than us. 

In \cite{Macpherson2019}, the authors used the averaging formalism introduced in \cite{Larena2009b}, considering averages on the simulation hypersurfaces but using kinematic fluid variables defined via the four-velocity $u^\alpha$ instead of the normal vector $n^\alpha$. In particular, the expansion rate was defined as $\theta \equiv h^{\alpha\beta} \nabla_\alpha u_\beta$, leading to different definitions of the scale factor, Hubble rate and backreaction terms considered here. Using that averaging scheme, the authors of \cite{Macpherson2019} found deviations in the matter and curvature density similar to our results, but they also found significantly larger deviations in the Hubble rate and scale factor as well as large backreaction\footnote{We note that \cite{Macpherson2019} plot the sum $\Omega_\mathcal{L}+\Omega_\mathcal{Q}$ different from the $\Omega_L$ and $\Omega_Q$ defined in this paper. $\Omega_\mathcal{L}+\Omega_\mathcal{Q}$ summarizes all the backreaction terms appearing in the equivalent of the Friedmann equation in \cite{Macpherson2019} and should therefore be compared to our $\Omega_Q$, which is the only additional term appearing in our Friedmann equation.}. The simulations used in \cite{Macpherson2019} had more small-scale structure than the simulations we consider and we cannot rule out that the significant backreaction found there is due to this small-scale structure. We however expect most of the difference in the two quantifications of backreaction (ours versus those presented in \cite{Macpherson2019}) to be due to the different averaging schemes, i.e. the different definitions of the fluid variables considered (ours in the $n^\alpha$ and theirs in the $u^\alpha$ frame).

In \cite{Williams2024}, the authors used the averaging formalism currently implemented in \texttt{mescaline} (different from ours), which is based on the averaging formalism introduced in \cite{Buchert2020}. The averages were used to compute expansion rates and curvature of voids, with the main focus of the paper being void statistics. The formalism presented in \cite{Buchert2020} considered averaging in domains comoving with the fluid flow, based on the kinematic fluid variables in the $u^\alpha$ frame. The currently implementation of the formalism used by \cite{Williams2024} approximates the fluid co-moving domains with domains co-moving with the hypersurfaces. The results we presented above indicate that this approximation might have to be considered carefully in future studies, since there is considerable mass in- and outflow for spheres comoving with the simulation spatial hyperaurfaces, meaning the spheres should deform to remain co-moving with the fluid and capture the correct expansion rate and scale factor. However, the results presented in \cite{Williams2024} are all at present time and hence the evolution of averages are not considered. We therefore do not expect that the results in \cite{Williams2024} are biased by this choice of hypersurface since the only computations affected by the (lack of) deformation of the spheres are those based on time derivatives. Computing the time derivatives in principle require considering simulation data and hence spheres at different time values, but as long as the time values are close together, the effect of matter in- and outflow would be small.
\newline\indent
Spatial averages and backreaction have also been studied in the context of \texttt{gevolution} simulations. In \cite{Adamek2018}, the authors considered averages both in domains co-moving with the fluid flow and on the hypersurfaces of the Newtonian gauge. The latter are (at least initially) the same as the ones considered here. The authors found negligible backreaction in the frame of the Newtonian hypersurfaces, with a relative deviation in the Hubble rate of at most $10^{-6}$ in simulation boxes of size $2048\;\mathrm{h}^{-1}\mathrm{Mpc}$ and $512\;\mathrm{h}^{-1}\mathrm{Mpc}$. The relative deviation in $H$ was slightly larger in $1/8$ sub-boxes, with a standard deviation of up to $10^{-4}$, but still small. The higher value in $H$ compared to our results could be explained by the fact that \texttt{gevolution} resolves more small-scale structure which is expected to source backreaction. This high resolution of structures is possible because \texttt{gevolution} is an N-Body simulation. In the co-moving frame, \cite{Adamek2018} found backreaction 3-5 orders of magnitude larger than in the Newtonian frame, but the authors note that they expect spatial averages on Newtonian gauge spatial hypersurfaces to be more relevant for observations, as we have also argued here. 

We also note that additionally to an EdS universe, \cite{Adamek2018} considers simulations with a cosmological constant $\Lambda$ more closely related to the real universe. This leads to a reduction of the backreaction at late times compared to the EdS case, as could be expected due to the accelerated expansion hindering the build up of structure. Since we only consider EdS simulations here, our results should be seen as upper bounds.

\section{Summary, Discussion and Conclusions}
\label{sec:5}
We introduced a spatial averaging scheme appropriate for practical implementation of averaging on the spatial hypersurfaces of Einstein Toolkit (ET) cosmological simulations. Using this averaging formalism, we studied the average evolution of sub-volumes in the form of spheres in three different ET simulations, considering 1000 spheres each of different radii in each simulation, confirming that our results are consistent between the simulations. We consistently find negligible backreaction in the spheres, even at small radii. This is in agreement with the results of \cite{Adamek2018} where backreaction was considered in \texttt{gevolution} simulations, in the sense that \cite{Adamek2018} also indicates negligible backreaction for this particular hypersurface choice. The backreaction found in \cite{Adamek2018} was, however, orders of magnitudes larger than that which we find. We attribute this to the higher resolution of small-scale structures available with N-body simulations. The promising work of \cite{Magnall2023} indicates that it may soon become possible to study small-scale structure formation also within ET cosmological simulations. It will be interesting to see how the stronger non-linearity and virialization of structures quantitatively affects the backreaction in this type of simulations. We also note that, on the face of it, our results do not agree with those of \cite{Macpherson2019} where significant backreaction was identified by averaging on the same hypersurfaces using other ET simulations. However, the results of \cite{Macpherson2019} were based on a different averaging scheme which is not readily comparable to ours, and in addition, the simulations used in \cite{Macpherson2019} contained higher resolution of structures.
\newline\indent
We find that there is significant in- and outflow of matter for the spheres on these hypersurfaces which suggests that it may be interesting to consider averages within volumes co-moving with the fluid flow using the mass preserving averaging formalism suggested in \cite{Buchert2020}. However, we note that the in- and outflow of matter and corresponding evolution of the curvature of spheres is countered by additional dynamical terms in the averaged acceleration equation. The resulting average (de-)acceleration of the spheres is thus very close to that of the entire simulation box.
\newline\indent
We argue that for spatial averages to be meaningful, they must be conducted on the hypersurfaces of statistical homogeneity and isotropy which for the considered simulations corresponds closely to the spatial simulation hypersurfaces, at least while in the linear regime. We base this claim on earlier studies indicating that in this case, spatial averages can be related sensibly to the main cosmological observables, the redshift and redshift-distance relation. Nonetheless, we cannot claim that our results indicate that cosmic backreaction is necessarily negligible in the real universe. For instance, it may be that other observables are more readily related to other spatial hypersurfaces. In such a case, assessing the relevance of backreaction for those observables will naturally require computing spatial averages on those hypersurfaces. Furthermore, future developments of the ET simulations that permit the development of structures on smaller scales may lead to a significantly larger backreaction. Lastly, we note that it is possible that the periodic boundary conditions of cosmological simulations artificially inhibit the evolution of backreaction, at least on global scales (see e.g. \cite{Coley2023, Coley2024, Buchert1997}). If this turns out to be the case, it is unclear to what extent such suppression propagates into sub-volumes and the quantifications of backreaction presented here.

\acknowledgments
The authors are funded by VILLUM FONDEN, grant VIL53032. The post-processing of the simulation data was performed using the UCloud interactive HPC system managed by the eScience Center at the University of Southern Denmark.
\newline\newline
The authors gratefully thank Hayley J. Macpherson for providing simulation data and \texttt{mescaline} as well as correspondence and comments on the manuscript. (The simulations used in this work were performed on the DiRAC@Durham facility managed by the Institute for Computational Cosmology on behalf of the STFC DiRAC HPC Facility (www.dirac.ac.uk). The equipment was funded by BEIS capital funding via STFC capital grants ST/P002293/1, ST/R002371/1 and ST/S002502/1, Durham University and STFC operations grant ST/R000832/1. DiRAC is part of the National e-Infrastructure.)
\newline\newline
{\bf Author contribution statement} 
AO conducted the analytical and numerical work under the guidance of SMK who also contributed with minor hands-on debugging. The original project-concept was developed by SMK but both authors contributed significantly to further developments of the project. Both authors contributed significantly to the writing of the paper.

\bibliography{main}

\widetext

\end{document}